\documentclass[onecolappendix]{aa}

\usepackage[varg]{txfonts}
\usepackage{graphicx,subfigure}
\usepackage{natbib}
\usepackage{color,bm}
\bibpunct{(}{)}{;}{a}{}{,} 

\newcommand{\beq}{\begin{equation}}
\newcommand{\eeq}{\end{equation}}
\newcommand{\beqn}{\begin{eqnarray}}
\newcommand{\eeqn}{\end{eqnarray}}
\renewcommand{\eqref}[1]{(\ref{#1})}

\newcommand{\pfrac}[2]{ \left(\dfrac{#1}{#2}\right) }
\newcommand{\bracket}[1]{\langle #1 \rangle}

\newcommand{\pd}{\partial}
\newcommand{\AU}{{\rm AU}} 
\newcommand{\kB}{k_{\rm B}}

\begin{document}

\title{On the water delivery to terrestrial embryos by ice pebble accretion}

\author{Takao Sato \and Satoshi Okuzumi \and Shigeru Ida}
\institute{
Department of Earth and Planetary Sciences, Tokyo Institute of Technology, Meguro, Tokyo, 152-8551, Japan\\
\email{okuzumi@geo.titech.ac.jp}
}
\abstract{
Standard accretion disk models suggest that the snow line in the solar nebula migrated interior to the Earth's orbit in a late stage of nebula evolution. In this late stage, a significant amount of ice could have been delivered to 1 AU from outer regions in the form of mm to dm-sized pebbles. This raises the question why the present Earth is so depleted of water (with the ocean mass being as small as 0.023~\% {of the} Earth mass). Here we quantify the amount of icy pebbles accreted by terrestrial embryos after the migration of the snow line {assuming that no mechanism halts the pebble flow in outer disk regions}. We use a simplified version of the coagulation equation to calculate the formation and radial inward drift of icy pebbles in a protoplanetary disk. The pebble accretion cross section of an embryo is calculated using analytic expressions presented by recent studies. We find that the final mass and water content of terrestrial embryos strongly depends on the radial extent of the gas disk, the strength of disk turbulence, and the time at which the snow lines arrives at 1 AU. The disk's radial extent sets the lifetime of the pebble flow, while turbulence determines the density of pebbles at the midplane where the embryos reside. We find that the final water content of the embryos falls below 0.023 wt\% only if the disk is compact (< 100 AU), turbulence is strong at 1 AU, and the snow line arrives at 1 AU later than 2--4 Myr after disk formation. If the solar nebula extended to 300 AU, initially rocky embryos would have evolved into icy planets of 1--10 Earth masses unless the snow-line migration was slow. If the proto-Earth contained water of $\sim 1$ wt\% as might be suggested by the density deficit of the Earth's outer core, the formation of the proto-Earth was possible with weaker turbulence and with earlier (> 0.5--2 Myr) snow-line migration.}

\maketitle

\titlerunning{Water delivery to terrestrial embryos by pebble accretion} \authorrunning{T.~Sato et al.}

\section{Introduction}\label{sec:intro}
Terrestrial planets in our solar system are characterized by their extremely low water content. 
The ocean of the Earth comprises only 0.023 wt\% of the total mass of the planet. 
The water content of the present Earth's interior is uncertain, 
but the 10\% density deficit of the Earth's outer core might suggest that 
that water of up to $\sim 1$ wt\% existed in the proto-Earth
and provided a large amount of hydrogen to the outer core (\citealt{O97,AOORD00}; 
see \citealt{NHU+14} for experiments supporting the large amount of hydrogen partitioning into the core).
An initial water content much in excess of $\sim 1~{\rm wt\%}$ seems unlikely 
{because neither stellar irradiation at $\sim 1~{\rm AU}$ \citep{MA10} 
nor giant impacts \citep{GA05} are able to vaporize the majority of the water 
from the Earth's gravitational potential.}
{Mars might possess, or might have possessed, subsurface water/ice 
of 0.01--0.1 \% of the total Mars mass \citep[][]{KSU+14}.
 Venus has a very dry atmosphere with the low-altitude ${\rm H_{2}O}$ mixing ratio 
 of 10--100 ppm \citep{PF87,DH92} and the high viscosity of Venus's mantle 
suggests that its interior is also dry \citep{NM98}.} 
The low water content of the terrestrial planets is in stark contrast 
to the high ice content of outer solar system bodies; 
in the extreme example of comets, the ice-to-rock mass ratio is generally thought to be close to unity
\citep[e.g.,][]{A11}. 

The fact that the Earth was born dry might not be surprising at first sight
given that the Earth's orbit is well inside the snow line of the current solar system.  
The snow line is defined by the orbit inside which water ice sublimates into vapor. 
Assuming that water ice sublimates at 170 K, 
the snow line of the present solar system lies at {about 3 AU} from the Sun.
{The snow line in the solar nebula, 
which is the protoplanetary disk that formed the solar system, 
would have been at the same location if the nebula was optically thin to direct stellar radiation \citep{H81}. }

However, {the solar nebula was presumably optically thick at least 
in its early evolutionary stage because of the presence of abundant small dust grains.
In an optically thick protoplanetary disk,
the snow line can be either inside or outside the Earth's orbit (1 AU) 
depending on how much accretion heating is effective.}
The disk is hottest in its earliest evolutionary stage  
where the central pre-main-sequence star is most luminous \citep[e.g.,][]{KNH70,TCCB12} 
and {where the accretion rate is the highest \citep[e.g.,][]{HCGD98}}. 
{Standard viscous accretion disk models show that 
the snow line, or the location where the gas temperature is 150--170 K, around a solar-mass star 
lies at about 5 AU when the accretion rate of the disk $\dot{M}$ is $10^{-7}~M_\sun~\rm yr^{-1}$
\citep[][]{D05,GL07,MDKD11,ONI11,BJLM15,BCP15,MCMP15}}. 
However, as the accretion rate decreases {with time}, 
{the disk cools down and the snow line moves inward. 
Assuming standard viscous accretion, the snow line passes 1 AU 
at $\dot{M} \approx$ a few $\times 10^{-9}~M_\sun~\rm yr^{-1}$
if all dust in the disk is in the form of opacity-dominating micron-sized grains
\citep{D05,GL07,ONI11,BJLM15},
and at $\dot{M} \approx 1\times 10^{-8}~M_\sun~\rm yr^{-1}$
if the grains are depleted by an order of magnitude \citep{BJLM15}. 
The migration of the snow line stops at $\sim 0.7~\AU$ (for the stellar luminosity of $1L_\odot$)
when accretion heating ceases to be important around these orbits \citep{SL00,D05,GL07,ONI11}.
At this stage, the interior of the disk is much colder than that of an optically thin disk 
because the disk can only receive stellar radiation on its surface.} 
The snow line moves out {toward} the final position $\approx 3$ AU
{only after} the interior of the disk becomes optically thin to direct stellar radiation \citep[$\dot{M} \la 10^{-10}~M_\sun~\rm yr^{-1}$,][]{ONI11}.

The above picture is based on {standard} accretion disk models 
{in which turbulence is assumed to be spatially uniform. }
For example, {accretion heating would be much less significant 
than anticipated by the uniformly turbulent models if the disk is only turbulent on its surface.
This is the case in magnetically driven accretion models where the magnetorotational instability \citep{BH91},
which is the driver of disk turbulence, is suppressed by magnetic diffusion near the midplane \citep{HT11,FFGC13}. } 
{On the other hand}, the nonsteady accretion model of \citet{ML12}, which 
incorporates the gravitational instability and suppression of magnetic turbulence by magnetic diffusion,
suggests that the gas temperature at 1 AU could be maintained high enough to sublimate ice  
even in the late stage of disk evolution.
{The snow line would not have reached the terrestrial region 
if X-ray-driven photoevaporation had cleared the gas in that region  
when $\dot{M} \approx 1\times 10^{-8}~M_\sun~\rm yr^{-1}$ \citep{OECA10}.}

The inward migration of the snow line, if it really occurs in  protoplanetary disks, 
gives two important constraints on the formation of terrestrial planets at $\sim 1~\AU$ like the Earth. 
{Earth-sized terrestrial planets are generally believed to form through giant impacts of 
Mars-sized ($\sim 0.1 M_\oplus$) solid bodies called planetary embryos \citep[e.g.,][]{WS89,KI02}}. 
{Given the inefficiency of removing water from embryos through giant impacts \citep{GA05}, 
water-devoid planets must form from water-devoid embryos.}
{Such embryos can form at 1 AU 
only when intense accretion heating pushes the snow line to $> 1~{\rm AU}$;
otherwise, like comets, they would have an ice-to-rock ratio of $\approx 1$.} 
{In the standard viscous disk models, this constraint means that the terrestrial embryos 
can only form when $\dot{M} \la$ a few to 10 $\times 10^{-9}~M_\sun~\rm yr^{-1}$
with the exact value depending on how much of the small dust grains are depleted (see above).
Assuming the correlation between the stellar age and mass accretion rate suggested by observations
\citep{HCGD98,BJLM15}, this also means that terrestrial planet formation needs to have been completed 
within $\sim$ 1--3 Myr after disk formation 
(there is, however, a large scatter in the $\dot{M}$--age correlation).
The Hf-W dating of Martian meteorites indicates that Mars, 
a possible planetary embryo that survived giant impacts, 
formed during the first $\sim$ 1--3 Myr of the solar system formation \citep{DP11}.
This Hf-W dating implies that the terrestrial embryo formation in the solar system barely satisfied this time constraint. 
One should keep in mind, however, that magnetically driven accretion models might predict very different 
time constraints as discussed above.}

The second important constraint is that the rocky embryos must avoid accretion of a significant amount of ice 
that could occur after the inward migration of the snow line. 
It is known that solid particles in a gas disk drift toward the central star because 
the gas drag robs the particles of angular momentum \citep{AHN76,W77a}.
The angular momentum loss is most effective  
for millimeter- to meter-sized particles that are marginally decoupled from the gas disk.   
Models incorporating the drift and coagulation of solid particles predict that
a significant amount of millimeter to decimeter-sized ice aggregates flow from outer disk regions
toward the snow line \citep[e.g.,][]{G07,BDH08,BDB10}. 
Without any mechanisms preventing the pebble flow, the total amount of ice that is delivered to
the inner orbits is comparable to the total amount of ice in the disk ($\sim 10$--$100M_\oplus$)
because the majority of solids in a disk generally reside in outer regions.  
The problem here is that large solid bodies like planetary embryos are efficient at capturing  
pebble-sized particles because of the help of the disk's gas drag \citep{OK10,LJ12}.
Therefore, if the snow line in the solar nebula migrated inside 1 AU, 
rocky embryos at 1 AU  could have accreted a non-negligible amount of ice.
{One might expect that this water delivery mechanism is 
potentially relevant to the origin of the Earth's ocean; 
however, the immediate problem with this interpretation is that 
the D/H ratio of icy pebbles from outer disk regions would presumably have been similar to those of comets,
which are on average higher than the Earth ocean water value \citep[e.g.,][]{ABB+15}. 
If this is the case, the amount of water supplied by the icy pebbles must have been much smaller 
than that of ocean water, or at least smaller than the water capacity of the Earth's interior,
in order to avoid an enhancement of the ocean D/H ratio.
} 

The question of how much water is delivered to terrestrial embryos by icy pebbles is closely linked to 
the so-called pebble accretion scenario for giant planet formation recently proposed by 
\citet[][{see also \citealt{KL14,JMLB15,MBC+16,MLJB15,MF15,LKD15,LKWB15}}]{LJ12,LJ14}.
They 
showed that efficient icy pebble accretion enables embryos of 1000 km in size outside the snow line 
to grow to the critical core mass for runway gas accretion  ($\sim 10~{M}_\oplus$) within the lifetime of protoplanetary disks. 
Our study focuses on another aspect of the pebble accretion scenario: 
while the radial pebble flux feeds giant planet cores in outer disk regions,  
the same pebble flux could deliver an excessive amount of water to terrestrial embryos in inner disk regions. 

{While this paper was under revision, a paper that discusses the issue of 
the snow-line migration appeared in print \citep{MBC+16}.
The paper proposes the scenario that proto-Jupiter halted the pebble flow from outer disk regions  
by carving a pebble-trapping gap in the gas nebula.
Although this is one plausible scenario (see also the discussions in our Sect.~\ref{sec:filter}),
it is also important to pursue the possibility that Earth-forming embryos avoided 
excessive water delivery even if no mechanism stopped the icy pebble flow.
This is the subject of this paper.
} 

{In this study,} we calculate the amount of ice accreted by an embryo at $\sim 1~\AU$ 
{based on the assumption that the snow line migrates inward across 1 AU.}
We employ a simple model of global dust evolution in which the collisional growth (coagulation) 
and radial drift of icy dust particles in a disk are treated in a self-consistent way. 
{We compute the amount of water delivered to a terrestrial embryo 
for a range of model parameters} including the strength of turbulence, 
the time at which the snow line moves interior to 1 AU, and the radial extent of the gas disk.
Our model is technically similar to the analytic model of \citet{LJ14} in that both treat 
the dominant particle size at each orbital radius instead of treating the full particle size distribution. 
An important difference from the previous study by \citet{LJ14} is that we apply 
the concept of pebble accretion to the problem of water delivery to terrestrial planets. 
In addition, our numerical model includes a more detailed calculation of the relative velocity 
between particles and also takes the finite radial extent of a protoplanetary disk into account, 
both of which affect the properties of the radial pebble flow. 
We also calibrate our model using the result of a detailed coagulation simulation that resolves 
the full particle size distribution \citep{OTKW12}.

The structure of the paper is as follows. 
In Sect.~\ref{sec:model}, we introduce the models of the solar nebula, dust evolution, and pebble accretion,
emphasizing how disk turbulence affects dust coagulation and pebble accretion quantitatively. 
In Sect.~\ref{sec:results}, we present the results of our model calculations and highlight 
how the rate of pebble accretion by an embryo depends on the radial extent 
of the disk and on turbulence strength.  
We give some discussions in Sect.~\ref{sec:discussion} and summarize in Sect.~\ref{sec:summary}.
Appendix~\ref{sec:simple} is devoted to the validation of the simplified dust evolution model employed in this work.

\section{Model}\label{sec:model}
\subsection{Overview}
\begin{figure}[t]
\centering
\includegraphics[width=9cm]{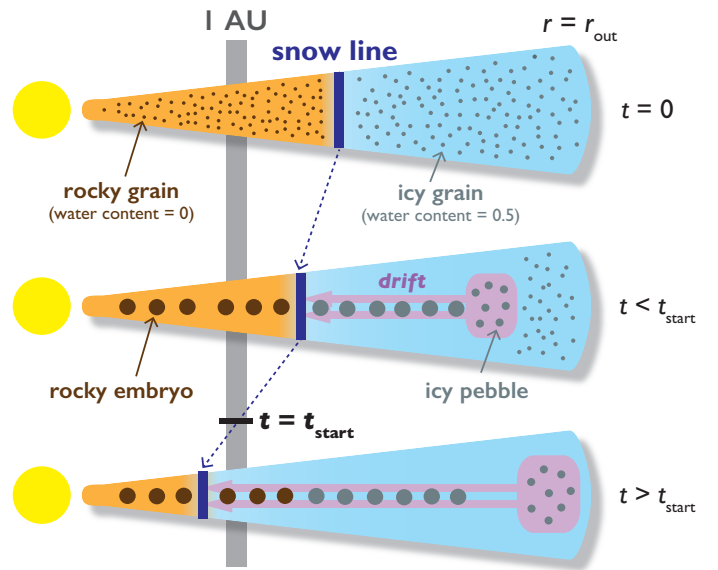}
\caption{
Schematic illustration showing the radial inward drift of icy pebbles 
and the inward migration of the snow line in a protoplanetary disk.  
Rocky embryos at 1 AU accrete radially drifting icy pebbles 
when the snow line resides at $< 1$ AU. 
Time $t = t_{\rm start}$, at which the snow line passes 1 AU, is taken as a free parameter.  
}
\label{fig:scenario}
\end{figure}

We describe the model we use to {quantify} how much water is delivered 
to rocky embryos at 1 AU through icy pebble accretion
(see Fig.~\ref{fig:scenario} for a schematic of the model).   
{We consider a solar-mass star and a protoplanetary disk of outer radius $r_{\rm out}$. } 
We assume that the snow line is initially well beyond 1 AU and migrates in across 1 AU
at time $t_{\rm start}$ after the beginning of dust evolution. 
We take $t_{\rm start}$ as a free parameter to avoid complications that would 
result from detailed modeling of the snow-line evolution.   
We assume that the solids in the disk 
are initially in the form of $0.1~{\rm \mu m}$-sized dust grains, 
and calculate the growth and radial inward drift of ice particles outside 
the snow line via a simplified dust coagulation model {described} in Sect.~\ref{sec:eqevol}.
The calculation gives us the mass flux (in the direction of the central star) and typical size 
of icy pebbles that arrive at 1 AU as a function of time $t$. 
{The ice-to-rock mass ratio of the icy pebbles is assumed to be unity
in accordance with the solar system composition compiled by \citet{L03}.
Millimeter observations of protoplanetary disks suggest that 
$r_{\rm out}$ is typically within the range $100$--$300~\rm AU$ \citep[e.g.,][]{AW07a}. 
We consider the two values $r_{\rm out} = 100~\AU$ and 300 AU. }

We then place a rocky embryo of initial mass $M_{\rm e,0}$ 
at 1 AU and allow it to accrete ice particles at $t > t_{\rm start}$. 
{We consider two cases, $M_{\rm e,0} = 10^{-1}M_\oplus$ and $10^{-2}M_\oplus$.
The larger $M_{\rm e,0}$ is the typical mass of terrestrial embryos predicted from 
planetesimal accumulation without fragmentation \citep[e.g.,][]{WS89,KI02}.
The final water fraction of the embryo is compared with the minimum water 
fraction of the present Earth given by the ocean (0.023 wt\%) and with 
the hypothetical water fraction of the proto-Earth inferred from 
the density deficit of the Earth's outer core  (1 wt\%).}

The radial drift of ice particles considered in this study is due to their angular momentum
in a sub-Keplerian rotating protoplanetary disk \citep{AHN76,W77a}. 
In reality, solids in a disk have an additional inward velocity owing to the accretion of the background 
gas onto the central star. 
However, this latter velocity component  is negligible compared to the former component  
whenever dust grows into pebble-sized particles \citep{BDH08,BKE12}.

We neglect possible filtration by planetesimals, planetary embryos, or gas giants exterior to 1 AU. 
As already shown by previous studies \citep{LJ14,GIO14,MN12},
a swarm of planetesimals or embryos filters only a minor fraction of the pebble flow (typically $< 50~\%$) 
unless {the size distribution of the bodies is narrowly peaked at $10^3$--$10^4$ km in radius} \citep[see][]{GIO14}.  
By contrast, if massive planets already exist at $t \sim t_{\rm start}$, they can efficiently 
{halt the flow of the pebbles} by opening a gap in the gas disk \citep[e.g.,][]{PM06,RAWL06,ZNDEH12,PBB12,MN12,LJM14}. 
By neglecting this effect, we effectively assume that such gap-forming planets 
form only after the snow line migrates inside 1 AU.    
We discuss this point in more detail in Sect.~\ref{sec:filter}.
We also neglect the loss of the pebble flux due to the accretion 
by adjacent rocky embryos.
Thus, the problem we are considering reduces to the problem of 
calculating the pebble accretion rate of each  isolated rocky embryo.

In the following subsections, we describe our disk model in Sect.~\ref{sec:disk}, 
the equations that determine the evolution of icy pebbles in Sect.~\ref{sec:eqevol} and \ref{sec:velocity},
our pebble accretion model in Sect.~\ref{sec:pebble}, and our parameter choices in Sect.~\ref{sec:param}.

\subsection{Disk model}\label{sec:disk}
The radial distribution of the gas surface density ${\Sigma}_{\rm g}$ is taken 
from the minimum mass solar nebula (MMSN) model of \citet{H81},
\beq
{\Sigma}_{\rm g} = 1700\left(\frac{r}{1\AU}\right)^{-3/2} {\rm g~cm^{-2}}, 
\label{eq:Sigmag}
\eeq
where $r$ is the distance from the central star. 
We cut off $\Sigma_{\rm g}$ at $r > r_{\rm out}$ 
and take the cutoff radius $r_{\rm out}$ as a free parameter (either 100 AU or 300 AU).
The initial dust surface density $\Sigma_{d,0}$ is taken to be $1\%$ of $\Sigma_{\rm g}$. 
The total dust mass within the initial disk is 
\beq
M_{\rm d} = 2\pi \int^{r_{\rm rout}}_{r_{\rm in}} r \Sigma_{\rm d,0} dr \approx 80M_\oplus \pfrac{r_{\rm out}}{100~{\rm AU}}^{1/2},
\label{eq:Md}
\eeq
where we have used $r_{\rm in} \ll r_{\rm out}$. 
Since $M_{\rm d}$ is an increasing function of $r_{\rm out}$, the dominant part of the mass 
resides in the outermost region of the disk. 
We come back to this point in Sect.~\ref{sec:global}.

The gas disk is assumed to be isothermal and hydrostatic in the vertical direction.
The gas density at the midplane is thus given by 
$\rho_{\rm g} = \Sigma_{\rm g}/(\sqrt{2\pi}h_{\rm g})$, 
where $h_{\rm g} = c_{\rm s}/\Omega$ is the gas scale height, 
$c_{\rm s} = \sqrt{k_{\rm B}T/m_{\rm g}}$ is the isothermal sound speed, 
$\Omega = \sqrt{GM_\sun/r^3} = 2.0\times 10^{-7}(r/1~\AU)^{-3/2}~{\rm s^{-1}}$
 is the Keplerian frequency with $k_{\rm B}$, $m_{\rm g}$, 
 $G$ being the Boltzmann constant, mean molecular mass (taken to be 2.34 amu),
and gravitational constant, respectively. 

As stated earlier, we do not directly treat the evolution of the snow line 
and instead express the migration of the snow line with $t_{\rm start}$. 
However, we do need a model of the gas temperature $T$ when we 
calculate the density structure of the gas disk and the thermal and turbulent velocity of particles.
For this purpose, we simply use a fixed power-law temperature profile
\begin{equation}
T = 170\left(\frac{r}{1\AU}\right)^{-1/2}~{\rm K},
\label{eq:T}
\end{equation}
where the slope has been taken from the {optically thin} disk model of \citet{H81}. 
{The value at $1~\AU$ has been chosen so that the snow line lies at $\sim 1~{\rm AU}$, 
which is motivated by 
our results that the water content of an embryo increases most rapidly just after the snow line 
passes the embryo ($t\approx t_{\rm start}$; see Sect.~\ref{sec:evol}). }
This temperature profile gives $c_{\rm s} = 7.8 \times 10^4(r/1~\AU)^{-1/4}~{\rm cm~s^{-1}}$, 
$h_{\rm g} = 0.026 (r/1~\AU)^{5/4}~{\rm AU}$, 
and $\rho_{\rm g} = 1.7\times 10^{-9} (r/1~\AU)^{-11/4}~{\rm g~cm^{-3}}$.
{In reality, in an optically thick disk, the radial temperature profile would be steeper 
than in Eq.~\eqref{eq:T} when accretion heating dominates \citep[e.g.,][]{LP80},
and would be shallower when stellar irradiation dominates \citep{KNH70,LP80}.
However, as we demonstrate in Sect.~\ref{sec:T}, the evolution and accretion of pebbles onto an embryo 
are fairly insensitive to the details of the temperature profile as long as an isolated single embryo is considered.
}

Our model takes the effects of disk turbulence on
the growth and vertical diffusion of dust particles  into account. 
{Turbulent diffusion is particularly important in our model 
because it determines the efficiency of pebble accretion by an embryo lying at the midplane 
\citep{GIO14,JMLB15,MLJB15,MF15}.}
We parametrize the turbulent diffusion coefficient as 
$D = \alpha c_{\rm s} h_{\rm g}$, where $\alpha$ is a dimensionless free parameter. 
If $D$ is equal to the turbulent viscosity (which is not used in this study), 
$\alpha$ corresponds to the viscosity parameter of \citet{SS73}.
For simplicity, we take $\alpha$ to be constant both in time and space.
The turnover time of the largest turbulent eddies, which is implicitly used in 
evaluating the turbulence-driven particle relative velocity, is taken to be $\Omega^{-1}$ 
in accordance with the typical correlation time of magnetorotational turbulence \citep[e.g.,][]{FP06}.
The role of disk turbulence as an effective viscosity is not taken into account in our model
since we do not evolve $\Sigma_{\rm g}$ or $T$.

\subsection{Dust growth and radial drift}\label{sec:eqevol}
We employ a simplified approach to calculate the mass flux of radially drifting pebbles.  
We assume that the mass distribution of dust particles at each orbit $r$ 
is singly peaked at a mass $m_{\rm p}(r)$. 
We then follow the evolution of the dust surface density $\Sigma_{\rm d}$ 
and peak mass $m_{\rm p}$ due to coagulation and radial drift by assuming 
that particles with mass $\sim m_{\rm p}$ dominate the dust surface density at each $r$.
Such a single-size approximation, also known as the two-moment bulk approximation in 
cloud modeling \citep{F94}, has been applied to modeling dust evolution in protoplanetary disks
\citep{KSR01,G07,BKE12} as well as in protoplanetary atmospheres \citep{O14}.\footnote{Recently, 
\cite{KODT16} proposed a single-size scheme based on the Lagrangian description.}
This allows us to track the global evolution of dust particles that dominate the radial mass flux 
at a much less computational cost than solving 
the exact coagulation equation that resolves the full particle size distribution. 
In Appendix~\ref{sec:simple}, we give analytic and numerical justifications 
of this approach as well as the formal definition of the peak mass $m_{\rm p}$.

Following \citet{BDH08}, the vertical distribution of the particles is approximated by a Gaussian 
$\propto \exp(-z^2/2h_d^2)$ and we determine the dust scale height $h_d$ from 
the balance between sedimentation and diffusion (see Eq.~\eqref{eq:hd} below). 
We neglect particle diffusion in the radial direction because 
its timescale ($\sim 10^4~{\rm yr}$ for $\alpha = 10^{-2}$ at 1 AU) 
is typically longer than the drift timescale of pebble-sized particles ($\sim 10^3~{\rm yr}$ at 1 AU). 

We assume that ice aggregates are so sticky that no fragmentation or bouncing occurs upon collision. 
{Numerical studies of aggregate collisions \citep{DT97,WTS+09,WTS+11,WTO+13} 
have shown that aggregates made of 0.1 $\mu \rm m$-sized icy grains undergo catastrophic 
disruption only at collision velocities higher than 50--80 ${\rm m~s^{-1}}$. 
With this high sticking efficiency, pebble-sized aggregates 
do not experience disruption in protoplanetary disks \citep[e.g.,][]{BDH08}.
\citet{KODT15} have recently pointed out that erosion by small projectiles 
might limit the growth of icy aggregates outside the snow line.
We do not consider this effect because the threshold velocity for erosion 
is still controversial (see the discussion in Sect. 2.3.2 of \citealt{KODT15}).
We also neglect condensation growth and sintering 
of icy aggregates in the vicinity of the snow line. 
While condensation would facilitates pebble growth \citep{RJ13}, 
sintering would induce destruction of pebble-sized aggregates \citep{S11a,S11b,OMSKT16}.
These processes could potentially change our results,
but we ignore them as a first step toward a full understanding of the problem.     
}
The aggregate internal density is fixed to be $\rho_{\rm int} = 1.4~{\rm g~cm^{-3}}$ for simplicity, 
but in reality the porosity of ice aggregates may evolve as they coagulate \citep{SWT08,OTKW12,KTOW13b}. 
Influences of the porosity evolution will be discussed in future work. 

Under the assumptions mentioned above, the equations governing the evolution of 
 $\Sigma_{\rm d}$ and $m_{\rm p}$ are given by 
\beq
\frac{\pd \Sigma_{\rm d}}{\pd t} + \frac{1}{r}\frac{\pd}{\pd r}(r v_{\rm r}\Sigma_{\rm d}) = 0,
\label{eq:evol_Sigmad}
\eeq
\beq
\frac{\pd m_{\rm p}}{\pd t} + v_{\rm r}\frac{\pd m_{\rm p}}{\pd r} 
= \frac{2\sqrt{\pi} a^2 \Delta v_{\rm pp}}{h_d}\Sigma_{\rm d},
\label{eq:evol_mstar}
\eeq
where $a = (3m_{\rm p}/4\pi \rho_{\rm int})^{1/3}$ is the particle radius corresponding to $m_{\rm p}$, 
and $v_{\rm r}$ and $\Delta v_{\rm pp}$ 
are the radial and relative velocities of the particles at the midplane, respectively. 
Our Eqs.~\eqref{eq:evol_Sigmad} and \eqref{eq:evol_mstar}
are essentially equivalent to Eqs.~(3) and (8) of \citet{O14}, although the deposition terms are not included here.
The formal derivation of Eqs.~\eqref{eq:evol_Sigmad} and \eqref{eq:evol_mstar}
from the exact coagulation equation is cumbersome (see Appendix~\ref{sec:moment}), 
but their physical interpretation is clear.  
Equation~\eqref{eq:evol_Sigmad} is merely the equation of continuity 
while Eq.~\eqref{eq:evol_mstar} states that the growth rate 
of peak-mass particles along their trajectory, 
$dm_{\rm p}/dt \equiv \pd m_{\rm p}/\pd t + v_{\rm r}\pd m_{\rm p}/\pd r $, 
is proportional to the product of the particle--particle collision cross section $\pi (a+a)^2 = 4\pi a^2$, 
relative velocity $\Delta v_{\rm pp}$, and dust density at the midplane $\Sigma_{\rm d}/(\sqrt{2\pi}h_{\rm d})$.\footnote{Precisely, the right-hand side of Eq.~\eqref{eq:evol_mstar} is $1/\sqrt{2}$ times  
the product of $4\pi a^2$, $\Delta v_{\rm pp}$, and  $\Sigma_{\rm d}/(\sqrt{2\pi}h_{\rm d})$.}

\subsection{Particle stopping time, scale height, and velocity}\label{sec:velocity}
The velocity and scale height of a particle depends on its stopping time $t_s$, 
which is the timescale of particle's momentum relaxation due to the gas drag.  
We evaluate $t_s$ with the piecewise formula
\begin{equation}
t_{\rm s} = \begin{cases}
\dfrac{\rho_{\rm int}a}{{\rho}_{\rm g}v_{\rm th}}, & a < \dfrac{9}{4}{\lambda}_{\rm mfp}, \\
\dfrac{4\rho_{\rm int}a^2}{9{\rho}_{\rm g}v_{\rm th}{\lambda}_{\rm mfp}}, & a > \dfrac{9}{4}{\lambda}_{\rm mfp},
\end{cases}
\label{eq:ts}
\end{equation}
where $v_{\rm th} = \sqrt{8\kB T/{\pi} m_{\rm g}}$ and ${\lambda}_{\rm mfp}$ are 
the thermal velocity and mean free path of gas particles, respectively. 
The mean free path is related to the gas density as 
${\lambda}_{\rm mfp} = m_{\rm g}/({\sigma}_{\rm mol}{\rho}_{\rm g})$, 
where ${\sigma}_{\rm mol} = 2.0 \times 10^{-15}~{\rm cm}^2$ is the molecular collision cross section.
The first and second expressions of Eq.~\eqref{eq:ts} are known as the Epstein and Stokes laws, respectively.
In many cases, it is useful to express the stopping time in terms of the dimensionless Stokes number 
\begin{equation}
{\rm St} \equiv {\Omega}t_{\rm s}.
\label{eq:St_def}
\end{equation}
Using the relations between $\Sigma_{\rm g}$, 
$\rho_{\rm g}$, $h_{\rm g}$, $\Omega$, $c_{\rm s}$, and $v_{\rm th}$, 
one can rewrite Eq.~\eqref{eq:ts} in terms of ${\rm St}$ as 
\beq
{\rm St} = \frac{\pi}{2}\frac{\rho_{\rm int} a}{\Sigma_{\rm g}}
\max\left(1, \frac{4a}{9\lambda_{\rm mfp}}\right).
\eeq

The radial drift velocity of particles is given by  \citep{AHN76,W77a}
\begin{equation}
v_{\rm r} = -\frac{2{\rm St}}{1 + {\rm St}^2}{\eta}v_{\rm K}, 
\label{eq:vr}
\end{equation} 
where 
\begin{equation}
\eta = - \frac{1}{2}\left(\frac{c_{\rm s}}{v_{\rm K}}\right)^2\frac{d\ln{(c_{\rm s}^2{\rho}_{\rm g})}}{d\ln{r}}
\label{eq:eta}
\end{equation}
is a dimensionless quantity characterizing the pressure gradient of the disk gas 
(which is the ultimate cause of the radial particle drift)
and $v_{\rm K} = r\Omega$ is the Kepler velocity. 
Our disk model gives $\eta = 1.1 \times 10^{-3} (r/1\AU)^{1/2}$ and 
${\eta}v_{\rm K} = 33~{\rm m~s^{-1}}$.
The value of $\eta$ is smaller than that of the optically thin MMSN model \citep{H81}
by the factor 0.6, which reflects the lower gas temperature in our model. 

The particle scale height is  given by \citep{DMS95,YL07,OTKW12}
\begin{equation}
h_{\rm d} = h_{\rm g}\left(1 + \frac{{\rm St}}{\alpha}\frac{1 + 2{\rm St}}{1 + {\rm St}}\right)^{-1/2}. 
\label{eq:hd}
\end{equation}
Equation~\eqref{eq:hd} assumes that vertical settling of the particles balances with 
vertical turbulent diffusion. 

The particle collision velocity $\Delta{v}_{\rm pp}$ is given by  
\begin{equation}
\Delta{v}_{\rm pp} 
= \sqrt{({\Delta}v_{\rm B})^2 + ({\Delta}v_{\rm r})^2 + ({\Delta}v_{\phi})^2 + ({\Delta}v_{\rm z})^2
+ ({\Delta}v_{\rm t})^2},
\label{eq:vpp}
\end{equation}
where $\Delta{v_{\rm B}}, \Delta{v_{\rm r}}, \Delta{v_{\phi}}, \Delta{v_{\rm z}}$ and $\Delta{v_{\rm t}}$ are
the relative velocities induced by Brownian motion, radial drift, azimuthal drift, vertical settling, and turbulence, 
respectively.
We evaluate these velocity components with the equations given in Sect~2.3.2 of \citet{OTKW12}, 
but with the assumption that the particle mass distribution is narrowly peaked at $m = m_{\rm p}$. 
For example, the Brownian component is given by $\Delta v_{\rm B} = \sqrt{8(m_1+m_2)\kB T/(\pi m_1m_2)}$,
where $m_1$ and $m_2$ are the masses of the colliding aggregates\footnote{There is a typographical error in the expression for $\Delta v_{\rm B}$ in \citet[][their Eq.~(17)]{OTKW12}.}, 
and we evaluate this by setting  $m_1 = m_2 = m_{\rm p}$. 
The differential drift velocities $\Delta v_{\rm r}$, $\Delta v_{\rm \phi}$, and $\Delta v_{\rm z}$
are functions of the Stokes numbers ${\rm St}_1$ and ${\rm St}_2$ of the colliding pair. 
Evaluation of these components within the single-size approximation 
is more tricky because they vanish for ${\rm St}_1 = {\rm St}_2$ but have a finite value 
for ${\rm St}_1 \sim {\rm St}_2$.
Since the real size distribution has a finite width,
the naive choice ${\rm St}_1 = {\rm St}_2 = {\rm St}(m_{\rm p})$
results in a significant underestimation of the particle velocities and, hence, of the particle growth rate
s shown in Appendix~\ref{sec:comparison}.
We introduce a dimensionless control parameter $\epsilon (<1)$ 
and set ${\rm St}_1 = {\rm St}(m_{\rm p})$ and ${\rm St}_2 = \epsilon{\rm St}(m_{\rm p})$
whenever we evaluate the drift velocities to take the effect of the size dispersion into account.
We  show in Appendix~\ref{sec:comparison} that the choice $\epsilon = 0.5$ 
best reproduces the results of a coagulation simulation that treats the fill size distribution.   
We adopt this choice throughout the paper. 
The turbulence-induced relative velocity $\Delta v_{\rm t}$ is also a function of ${\rm St}_1$
and ${\rm St}_2$ (assuming that the turnover time of the largest turbulent eddies is $\Omega^{-1}$), 
and we evaluate it using Eqs.~(16)--(18) of \citet{OC07}.
For ${\rm St}_1 \sim {\rm St}_2 \ll 1$, which is true for pebble-sized particles, 
$\Delta v_{\rm t}$ has approximate expressions (see Eqs.~(27) and (28) of \citealt{OC07}) 
\beq
{\Delta}v_{\rm t} \approx \begin{cases}
\sqrt{\alpha}c_{\rm s}{\rm Re}_{\rm t}^{1/4}|{\rm St}_1 - {\rm St}_2|, 
& {\rm St}_1 \ll {\rm Re}_{\rm t}^{-1/2}, \\[2mm]
\sqrt{3\alpha}c_{\rm s} {\rm St}_1^{1/2}, & {\rm St}_1 \gg {\rm Re}_{\rm t}^{-1/2},
\end{cases}
\label{eq:vcolt}
\eeq
where ${\rm Re}_{\rm t} = D/\nu_{\rm mol}$ is the turbulent Reynolds number 
and $\nu_{\rm mol} = v_{\rm th}{\lambda}_{\rm mfp}/2$ is the molecular viscosity. 
We  set ${\rm St}_1 = {\rm St}(m_{\rm p})$ and ${\rm St}_2 =  \epsilon{\rm St}(m_{\rm p})$
when evaluating $\Delta v_{\rm t}$ since it vanishes for ${\rm St}_1 = {\rm St}_2$ 
in the case of ${\rm St}_1 \ll {\rm Re}_{\rm t}^{-1/2}$. 

\begin{figure}[t]
\centering
\includegraphics[width=9cm]{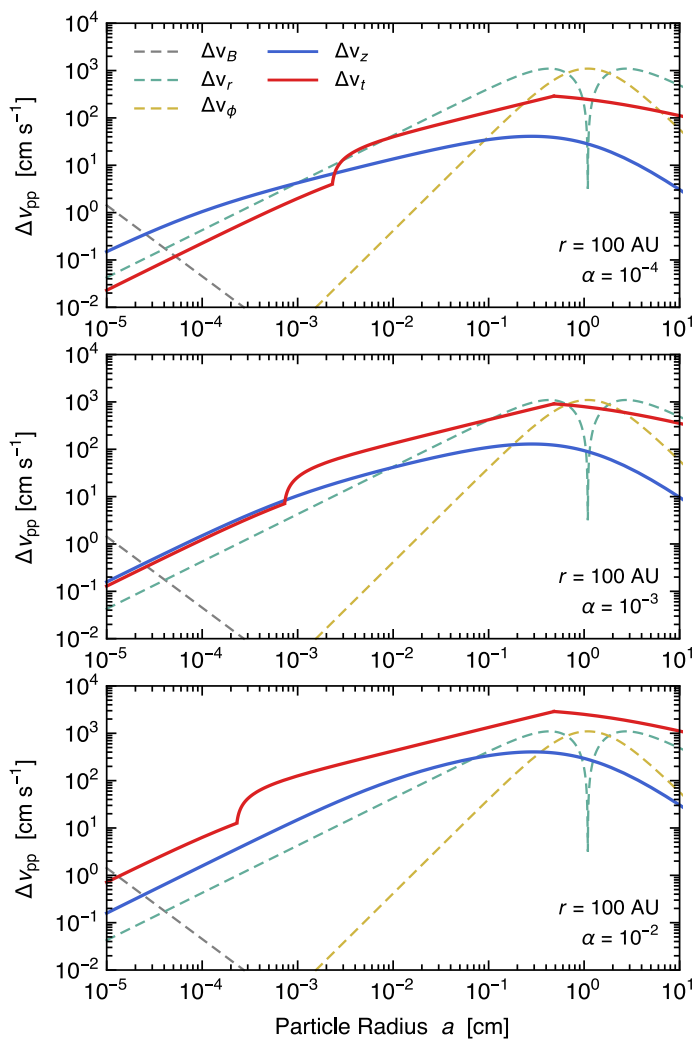}
\caption{Components of the particle relative velocity $\Delta v_{\rm pp}$ at $100~{\rm AU}$
as a function of particle radius $a$ for three values of the turbulence parameter 
$\alpha = 10^{-4}$ (top panel), $10^{-3}$ (middle panel), and $10^{-2}$ (bottom panel).
The velocity components that depend on $\alpha$ are shown by the solid curves.
The stopping time ratio of $\epsilon = 0.5$ is assumed 
for $\Delta v_{\rm r}$, $\Delta v_\phi$, $\Delta v_{\rm z}$, and $\Delta v_{\rm t}$.
}
\label{fig:dv}
\end{figure}
Which component of $\Delta v_{\rm pp}$ dominates depends not only on the particle size 
but also on the turbulence strength.  
To illustrate this, in Fig.~\ref{fig:dv} we plot all components of $\Delta v_{\rm pp}$ 
at $r = 100~{\rm AU}$ as a function of $a$ for different values of $\alpha$.
We assume $\epsilon=0.5$ when evaluating the non-Brownian components.
In general, the particle relative velocity has a maximum at ${\rm St} \approx 1$, 
which corresponds to $a \approx 1~{\rm cm}$ at this location. 
When $\alpha = 10^{-4}$, laminar components such as 
$\Delta v_{\rm z}$ and $\Delta v_{\rm r}$ are dominant for all $a$.
The turbulent component $\Delta v_{\rm t}$ becomes more important when $\alpha = 10^{-3}$, 
and dominates over the laminar components for all $a$ when $\alpha = 10^{-2}$. 

\begin{figure}[t]
\centering
\includegraphics[width=9cm]{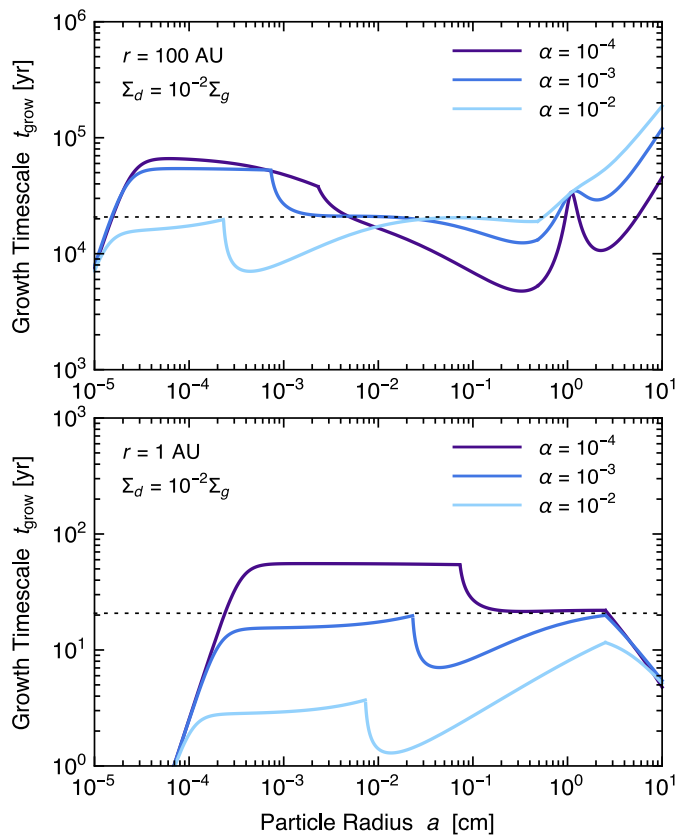}
\caption{Particle growth timescale $t_{\rm grow}$ (Eq.~\eqref{eq:tgrow}) 
at $100~{\rm AU}$ as a function of particle radius $a$ 
for $\Sigma_{\rm d} = 10^{-2}\Sigma_{\rm g}$ with 
different values of the turbulence parameter $\alpha$.
The dotted line shows the simple estimate 
$t_{\rm grow} = (4/\sqrt{3\pi})(\Sigma_{\rm g}/\Sigma_{\rm d})\Omega^{-1}$ (Eq.~\eqref{eq:tgrow_approx}).
}
\label{fig:tgrow}
\end{figure}
Since $\Delta v$ and $h_{\rm d}$ are already given,  
we are able to estimate the timescale of dust growth as a function of 
particle size. 
Here we define the particle growth timescale as 
\beq
t_{\rm grow} \equiv  \left(\frac{1}{a}\frac{da}{dt} \right)^{-1} 
=3 \left(\frac{1}{m_{\rm p}}\frac{dm_{\rm p}}{dt} \right)^{-1} 
= \frac{3m_{\rm p}h_{\rm d}}{2\sqrt{\pi} a^2 \Delta v_{\rm pp} \Sigma_{\rm d}},
\label{eq:tgrow}
\eeq
where $(d/dt) \equiv (\pd/\pd t) + v_{\rm r}(\pd/\pd r)$ 
is the Lagrangian time derivative, and we have used Eq.~\eqref{eq:evol_mstar} in the final expression. 
Figure~\ref{fig:tgrow} shows $t_{\rm grow}$ at 100~AU and 1~AU 
as a function of $a$ for three different values of $\alpha$.
Here, the dust-to-gas ratio is taken to be the initial value $\Sigma_{\rm d,0}/\Sigma_{\rm g} = 10^{-2}$.
It can be seen that $t_{\rm grow} \sim 10^4$--$10^5$ yr at 100~AU and $\sim 1$--$100$ yr at 1~AU,
indicating that $t_{\rm grow}$ scales approximately linearly with the orbital timescale 
$\propto \Omega^{-1} \propto r^{3/2}$. 
In fact, one can show that the simple relation 
\begin{align}
t_{\rm grow} &\approx \frac{4}{\sqrt{3\pi}}\frac{\Sigma_{\rm g}}{\Sigma_{\rm d}}\Omega^{-1}
\approx 2\times 10^4\pfrac{\Sigma_{\rm g}/\Sigma_{\rm d}}{100}\pfrac{r}{100~{\rm AU}}^{3/2}~{\rm yr}
\label{eq:tgrow_approx} 
\end{align}
holds in the special case where the conditions $a \ll \lambda_{\rm mfp}$, 
$\Delta v_{\rm pp} \approx \Delta v_{\rm t}$, and 
$ {\rm St}_1 \gg \max( {\rm Re}_{\rm t}^{-1/2}, \alpha)$ are met
(see \citealt{TL05} and \citealt{BDH08} for the derivation).
This expression, which is employed in the analytic pebble formation model of \citet{LJ14}, 
is extremely useful as an order-of-magnitude estimate, 
since it only depends on the gas-to-dust ratio and orbital frequency. 
However, caution should be exercised when using Eq.~\eqref{eq:tgrow_approx} 
for more precise calculations because the expression is less accurate 
if one or more of the conditions mentioned above is not met. 
For example, we can see in Fig.~\ref{fig:tgrow} that Eq.~\eqref{eq:tgrow_approx} 
overestimates the actual growth timescale 
by a factor of several when $\alpha \leq 10^{-3}$ and $a \la 10^{-3}~{\rm cm}$, 
for which $\Delta v_{\rm pp} \approx \Delta v_{\rm z}$ (see Fig.~\ref{fig:dv}) 
and ${\rm St} \la \alpha$.
For this reason, the time required for micron-sized dust particles to grow into pebbles 
is longer when $\alpha \leq 10^{-3}$ than when $\alpha = 10^{-2}$.

\subsection{Pebble accretion}\label{sec:pebble}
\begin{figure}[t]
\centering
\includegraphics[width=9cm]{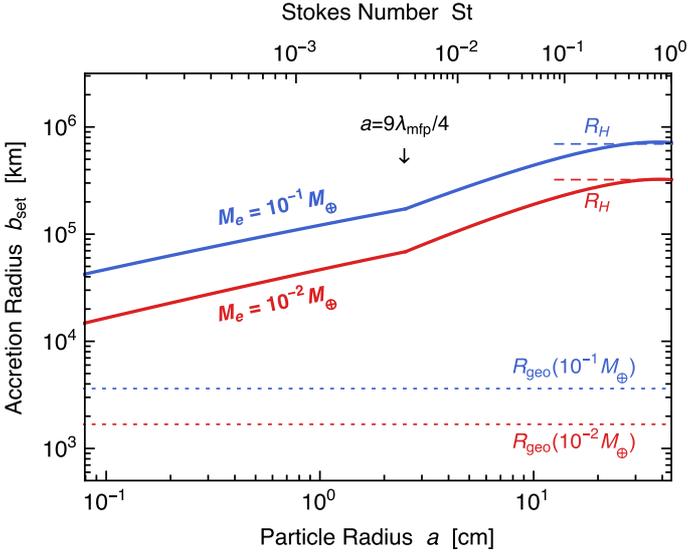}
\caption{Pebble accretion radius of an embryo $b_{\rm set}$ (Eq.~\eqref{eq:bset}; solid curve)
as a function of the pebble radius $a$. 
The upper and lower solid curves show $b_{\rm set}$ for embryos of masses  
$M_{\rm e} = 10^{-1}M_\oplus$ and $10^{-2}M_\oplus$, respectively.  
The dashed and dotted lines indicate the Hill radii $R_{\rm H}$ (Eq.~\eqref{eq:RH})
and geometric radii $R_{\rm geo} = (3M_{\rm e}/4\pi \rho_{\rm e})^{1/3}$ of the embryos,
respectively, where we take $\rho_{\rm e} = 3~{\rm g~cm^{-3}}$.
}
\label{fig:b}
\end{figure}
\begin{figure}[t]
\centering
\includegraphics[width=9cm]{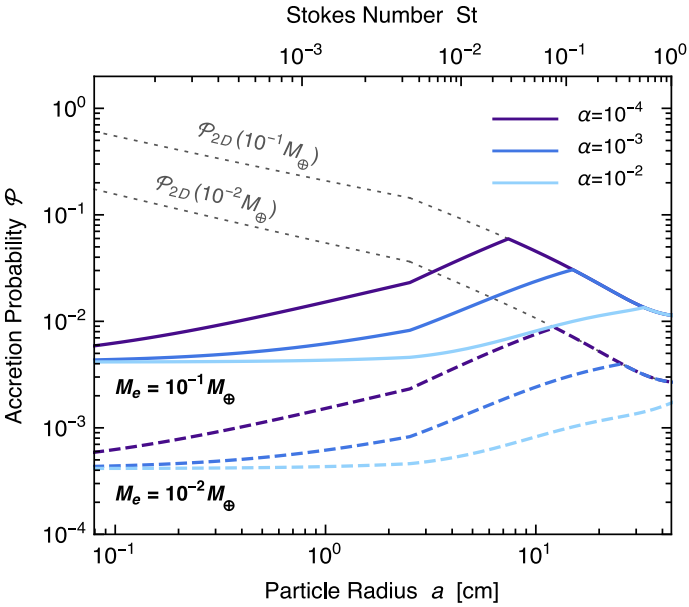}
\caption{Pebble accretion probability by a single embryo, ${\cal P}$ (Eq.~\eqref{eq:P}),
as a function of the pebble radius $a$ for different values of the turbulence parameter $\alpha$. 
The solid and dashed curves are for embryos of masses $M_{\rm e} = 10^{-1}M_\oplus$ 
and $10^{-2}M_\oplus$, respectively. The dotted lines indicate the accretion probability 
in the two-dimensional limit, ${\cal P}_{\rm 2D} = 2b_{\rm set}\Delta v_{\rm set}/(2\pi r|v_{\rm r}|)$. 
}
\label{fig:P}
\end{figure}

As mentioned at the beginning of Sect.~\ref{sec:model}, 
we place an rocky embryo at 1~AU in a protoplanetary disk 
and allow it to accrete icy pebbles at times $t > t_{\rm start}$. 
Following \citet{GIO14}, we evaluate the rate of pebble accretion 
by an embryo, $\dot{M}_{\rm e}$, as 
\beq
\dot{M}_{\rm e} = \min\left(2b_{\rm set}, \frac{\pi b_{\rm set}^2}{\sqrt{2\pi}h_{\rm d}}\right)
\Delta v_{\rm set}\Sigma_{\rm d},
\label{eq:Medot}
\eeq
where $b_{\rm set}$ is the effective pebble accretion radius of the embryo and 
\beq
\Delta v_{\rm set} = \eta v_{\rm K} + \frac{3}{2}b_{\rm set} \Omega
\label{eq:Deltav_set}
\eeq
is the (maximum) encounter velocity of the embryo and pebbles.
In Eq.~\eqref{eq:Medot}, the factor $\min(\cdots)$ accounts for the effect 
of the sedimentation of the pebbles onto the midplane: 
the accretion is two-dimensional 
($\dot{M}_{\rm e} \approx 2b_{\rm set}\Delta v_{\rm set}\Sigma_{\rm d}$) 
for $b_{\rm set} \gg h_{\rm d}$ and three-dimensional 
($\dot{M}_{\rm e} \approx \pi b_{\rm set}^2\Delta v_{\rm set}\rho_{\rm d}$,
where $\rho_{\rm d} = \Sigma_{\rm d}/\sqrt{2\pi}h_{\rm d}$ is the midplane pebble density)
for the opposite limit \citep[see Sect.~3.2 of][]{GIO14}. 
In the 3D case, $\dot{M}_{\rm e} (\propto h_d^{-1})$ decreases with increasing $\alpha$,
reflecting the fact that turbulence diffuses pebbles away from the midplane where the embryo resides.  
The first and second terms on the right-hand side of Eq.~\eqref{eq:Deltav_set}
represents the encounter velocity arising from 
the sub-Keplerian orbital velocity of the pebbles and from the Keplerian shear, respectively 
(see Sect.~5.1.3 of \citealt{OK10}). Equation~\eqref{eq:Deltav_set} neglects
the turbulence-driven encounter velocity $\approx \sqrt{\alpha} c_{\rm s}$, 
but this does not affect our results significantly as long as $\alpha \la 10^{-3}$
(for which $\sqrt{\alpha}c_{\rm s} \la \eta v_{\rm K}$). 

For the accretion radius $b_{\rm set}$, 
we use a simple empirical relation \citep{OK12} 
\beq
b_{\rm set} = b_{\rm set,0} \exp\left[-({\rm St}/2)^{0.65} \right].
\label{eq:bset}
\eeq
Here, $b_{\rm set,0}$ is  the solution to \citep{OK10}
\beq
3b_{\rm set,0}^3 + 2\eta r b_{\rm set,0}^2 - 24{\rm St}R_{\rm H}^3 = 0,
\label{eq:bset0}
\eeq
where
\beq
R_{\rm H} \equiv \pfrac{M_{\rm e}}{3M_*}^{1/3}r
\label{eq:RH}
\eeq
is the Hill radius of the embryo. 
Equations~\eqref{eq:bset} and \eqref{eq:bset0} apply 
when the particles are coupled to the disk gas so strongly that they 
accrete onto the embryo at a terminal velocity.
\citet{OK10} called this regime the {\it settling regime}.
Equation~\eqref{eq:bset0} originally comes from the consideration  
that an embryo accretes pebbles whose trajectories are greatly deflected by the embryo's gravity \citep{OK10,LJ12}.
The exponential cutoff appearing in Eq.~\eqref{eq:bset} 
assumes that $R_{\rm H}$ is considerably smaller than
the Bondi radius of the embryo defined with respect to the headwind $\eta v_{\rm K}$,
\beq
R_{\rm B} \equiv \frac{GM_{\rm e}}{(\eta v_{\rm K})^2}.
\label{eq:RB}
\eeq
{Since $(R_{\rm H}/R_{\rm B})^3 = (\eta^6/3)(M_{\rm e}/M_*)^{-2}$, 
the assumption $R_{\rm H} < R_{\rm B}$ holds when $M_{\rm e} > 0.58 \eta^3 M_*$.  
At 1 AU, this condition is satisfied for  $M_{\rm e} \ga 10^{-3}~ M_\oplus$.
The exponential cutoff accounts for the fact that efficient accretion through settling takes place only 
when ${\rm St} \ll 1$. 
The Bondi and Hill radii satisfy the relations $R_{\rm B}/(\eta r) = M_{\rm e}/(\eta^3 M_*)$ 
and $R_{\rm H}/(\eta r) = 0.58 (R_{\rm B}/R_{\rm H})^{1/2}$. 
The latter relation implies that $R_{\rm H}  \gg \eta r$ when $R_{\rm H} \ll R_{\rm B}$. }

As pointed out by \citet{LJ12} and \citet{GIO14}, the settling regime can be divided into two subregimes depending on 
which of the headwind $\eta v_{\rm K}$ and Keplerian shear $3b_{\rm set}\Omega/2$ 
dominates. When $\eta v_{\rm K} \gg 3b_{\rm set}\Omega/2$,
or equivalently $b_{\rm set} \ll 2\eta r/3$, the first term 
in Eq.~\eqref{eq:bset0} is negligible compared to the second term, 
and hence $b_{\rm set}$ is approximately given by 
\beq
b_{\rm set} \approx \sqrt{\frac{12{\rm St}R_{\rm H}^3}{\eta r}} 
= 2\sqrt{\frac{GM_{\rm e}t_{\rm s}}{\eta v_{\rm K}}}.
\label{eq:bset0_B}
\eeq
This regime was referred to as the drift accretion regime by \citet[][see their Eq.~(27)]{LJ12} 
and the Bondi regime by \citet[][the third expression of their Eq.~(C.3)]{GIO14}. 
In the opposite limit, $b_{\rm set} \gg 2\eta r/3$,
the second term in Eq.~\eqref{eq:bset0} is negligible, and we obtain
\beq
b_{\rm set} \approx 2{\rm St}^{1/3} R_{\rm H}
\label{eq:bset0_H}
\eeq
{($b_{\rm set} < R_{\rm H}$ because ${\rm St} \ll 1$ in the settling regime).} 
This corresponds to the Hill accretion regime of \citet[][see their Eq.~(40)]{LJ12}. 
This regime is also essentially equivalent to the Hill regime of \citet{GIO14}, 
but the factor $2{\rm St}^{1/3}$ appearing in our Eq.~\eqref{eq:bset0_H} 
is neglected in their limiting expression
for $b_{\rm set}$ (the fourth expression of their Eq.~(C.3)).  
A comparison between Eq.~\eqref{eq:bset0_B} and Eq.~\eqref{eq:bset0_H}
shows that the Hill accretion applies (i.e., $\eta v_{\rm K} \gg b_{\rm set} \Omega$) 
when ${\rm St}$ satisfies 
\beq
{\rm St} 
 = \frac{\eta^3 M_*}{9M_{\rm e}} 
\approx  4\times 10^{-4} \pfrac{\eta}{10^{-3}}^3 \pfrac{M_{\rm e}}{10^{-1}~M_\oplus}^{-1}\pfrac{M_*}{M_\sun}.
\label{eq:St_bnd}
\eeq
At 1 AU, this condition is equivalent to $a \ga 3~{\rm mm}$. 
As we see in Sect.~\ref{sec:1AU}, the pebbles drifting to 1 AU mostly satisfy this condition.

As an example, Fig.~\ref{fig:b} shows $b_{\rm set}$ of an embryo located at 1 AU
as a function of the pebble size $a$ and of the pebble Stokes number ${\rm St}$.
For comparison, we also plot the embryo's geometric radius 
$R_{\rm geo} = (3M_{\rm e}/4\pi \rho_{\rm e})^{1/3}$ where the embryo's internal density 
$\rho_{\rm e}$ is set to $3~{\rm g~cm^{-3}}$. 
For $M_{\rm e} \geq 10^{-3} M_\oplus$, the accretion radius is considerably larger than $R_{\rm geo}$ 
as long as $a \ga 0.1~{\rm mm}$ \citep[see also Figure~10 of][]{GIO14}. 
At ${\rm St} \sim 1$ ($a \sim 1~{\rm m}$), 
the accretion radius reaches the Hill radius
$R_{\rm H} = (4\pi \rho_{\rm e} r^3/9M_*)^{1/3} R_{\rm geo} \approx 200 R_{\rm geo}$.
\begin{table*}
\centering
\begin{tabular}{lll}
\hline\hline
Quantity & Description & Value \\ \hline
$r_{\rm out}$ (AU) & Disk size & 100, 300 \\
$\alpha$ & Turbulence parameter  & $10^{-2}$, $ 10^{-3}$, $10^{-4}$\\
$M_{\rm e,0}$ $(M_\oplus)$ & Embryo mass before pebble accretion & $10^{-1}$, $10^{-2}$ \\ 
$t_{\rm start}$ (Myr) & Initial time of pebble accretion &  0.5, 1, 2, 3, 4, 5\\ 
\hline
\end{tabular}
\caption{List of free parameters in this study.}
\label{tb:parameters}
\end{table*}

Fig.~\ref{fig:P} shows some examples of the pebble accretion rate as a function of $a$. 
We here express the accretion rate in terms of the accretion probability 
\beq
{\cal P} \equiv \frac{\dot{M}_{\rm e}}{\dot{M}_{\rm r}}
= \frac{\min\left(2b_{\rm set}, \frac{\pi b_{\rm set}^2}{\sqrt{2\pi}h_{\rm d}}\right)
\Delta v_{\rm set}}{2\pi r|v_{\rm r}|},
\label{eq:P}
\eeq
where 
\beq
\dot{M}_{\rm r} \equiv 2\pi r |v_{\rm r}| \Sigma_{\rm d}
\label{eq:Mdotr}
\eeq
is the radial inward mass flux 
of dust in the gas disk. 
By construction, ${\cal P}$ measures the fraction of radially drifting pebbles 
that are filtered by a single embryo.  
The accretion probability depends on the turbulence strength $\alpha$ 
via the pebble scale height $h_{\rm d}$. 
For comparison, the accretion probability in the two-dimensional limit,
${\cal P}_{\rm 2D} \equiv  2b_{\rm set} \Delta v_{\rm set}/(2\pi r|v_{\rm r}|)$, is also plotted. 
Since $h_{\rm d}$ decreases with increasing $a$, accretion of large particles 
(typically of sizes $a \ga 10~{\rm cm}$) takes place in a 2D manner. 
In this case, the accretion probability decreases with increasing $a$
because larger particles have a higher drift speed $|v_{\rm r}|$ 
($b_{\rm set}$ and $\Delta v_{\rm set}$ also increase with $a$, 
but more slowly than $|v_{\rm set}|$). 
Accretion of smaller particles ($a \la 10~{\rm cm}$) is limited by their large scale height 
$h_{\rm d}$ compared to the accretion radius $b_{\rm set}$. 
For these reasons, the accretion probability has a maximum at the pebble size 
corresponding to $h_{\rm d} \approx b_{\rm set}$.
The maximum probability is $\sim 10^{-2}$--$10^{-1}$ for $M_{\rm e} = 10^{-1}M_\oplus$ 
and $\sim 10^{-3}$--$10^{-2}$ for $M_{\rm e} = 10^{-2}M_\oplus$.

\subsection{Parameter choice}\label{sec:param}
The free parameters of our model are 
the disk size $r_{\rm out}$, turbulence parameter $\alpha$, 
embryos mass before icy pebble accretion $M_{\rm e,0}$, and 
the initial time $t_{\rm start}$ of icy pebble accretion by an embryo at 1~AU.  
Table~\ref{tb:parameters} lists the parameter choices adopted in this study.

\begin{figure*}
\centering
\includegraphics[width=17cm]{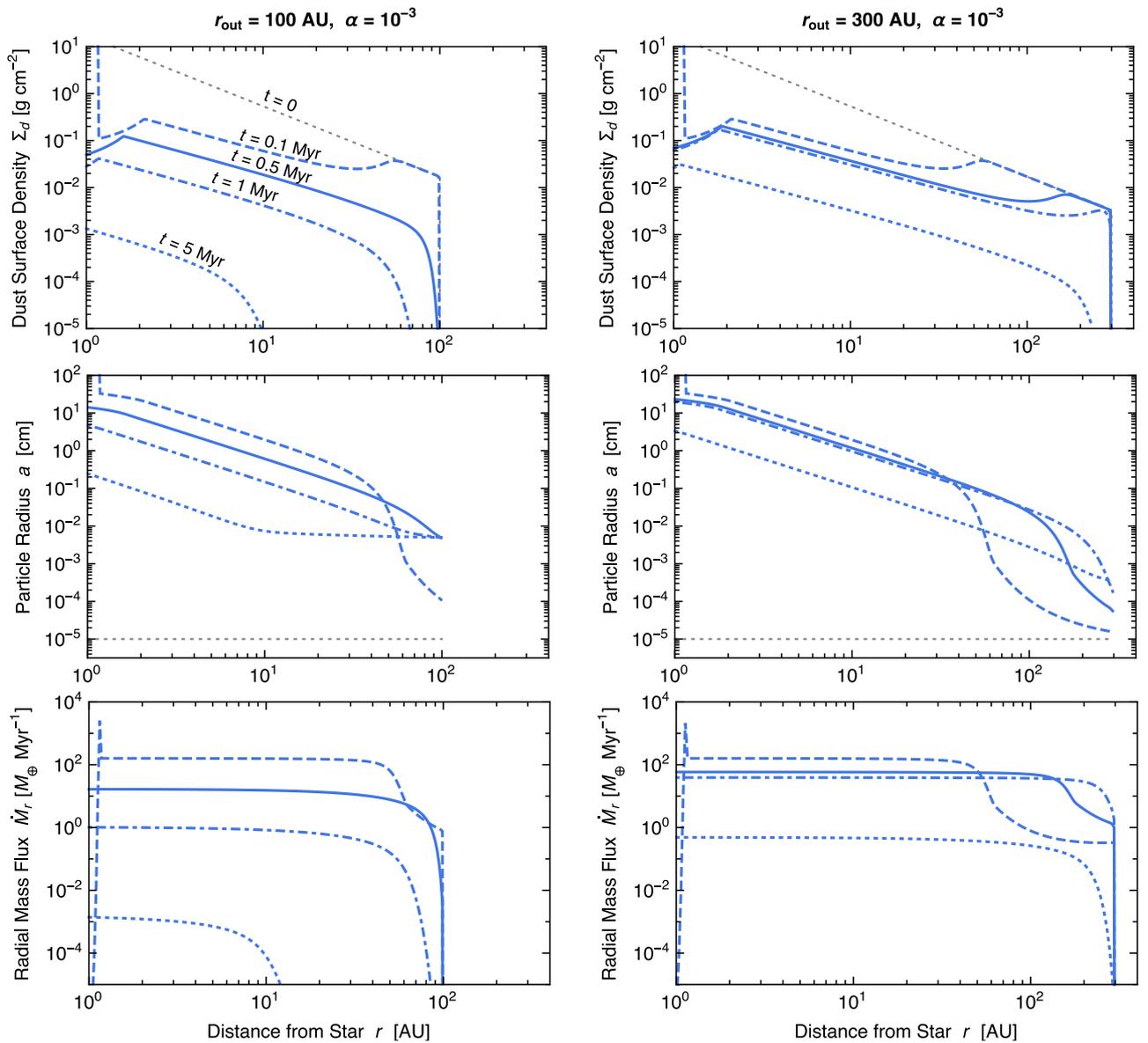}
\caption{Time evolution of the surface density $\Sigma_{\rm d}$ (top panels),
radius $a$ (middle panels), and radial mass flux $\dot{M}_{\rm r}$ (bottom panels) of dust particles 
as a function of orbital radius $r$ for models with $\alpha = 10^{-3}$ and 
with $r_{\rm out} =$ 100 AU (left panels) and 300 AU (right panels).
The black dotted lines show the initial condition, while the blue dashed, solid, 
dash-dotted, and dotted lines are the snapshots at times $t =$ 0.1, 0.5, 1, and 5 Myr, respectively.   
The jumps in $\Sigma_{\rm d}$ and $a$ at $r \approx 1~{\rm AU}$, $t = 0.1~{\rm Myr}$
are caused by rapid coagulation of particles initially located at these orbits  (see text). 
}
\label{fig:dr}
\end{figure*}

\section{Results}\label{sec:results}
We now present the results of our calculations step by step. 
In Sect.~\ref{sec:global}, we start by presenting the global evolution of icy particles  
to emphasize how the parameters $r_{\rm out}$ and $\alpha$ {control} the lifetime of 
the radial inflow of icy pebbles in a disk.
Sect.~\ref{sec:1AU} presents the properties of drifting pebbles at 1~AU in more detail. 
We then consider a rocky embryo forming at the Earth's orbit and calculate 
its pebble accretion rate in Sect.~\ref{sec:Medot}. 
The resulting evolution of the embryo's mass and water content is presented in 
Sect.~\ref{sec:evol}.

\subsection{Global picture of dust evolution}\label{sec:global}
\begin{figure}[t]
\centering
\includegraphics[width=9cm]{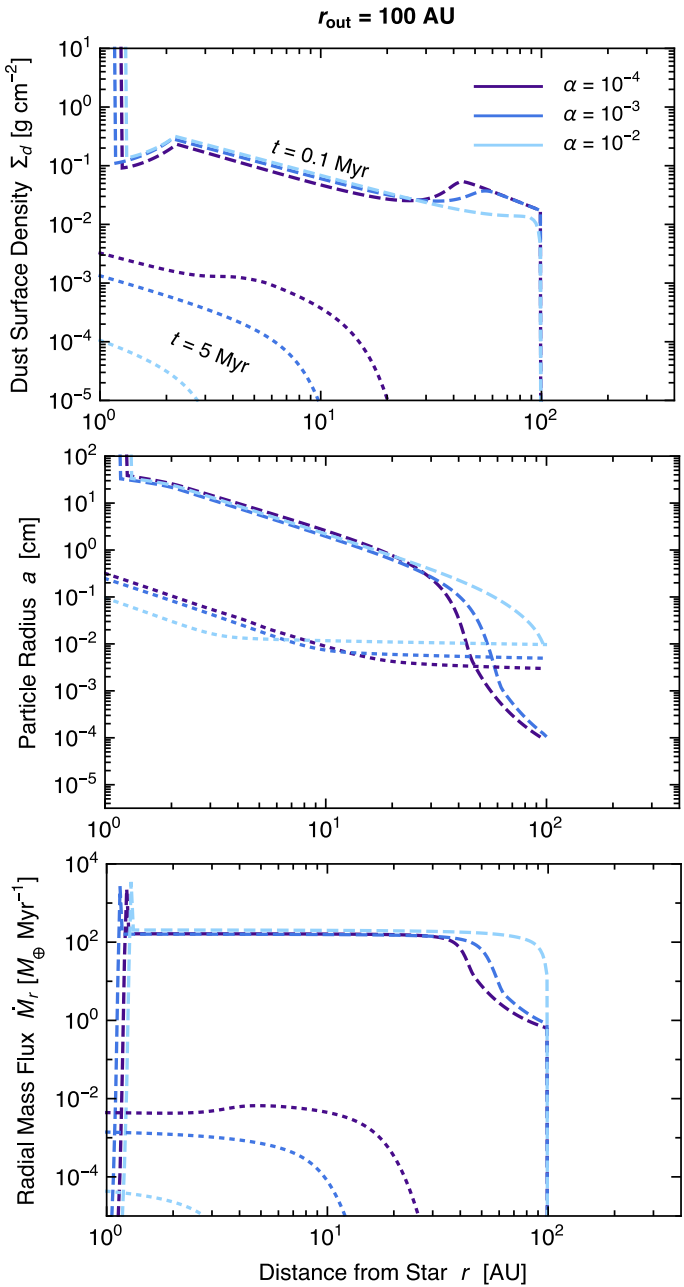}
\caption{Time evolution of the surface density $\Sigma_{\rm d}$ (top panel),
radius $a$ (middle panel), and radial mass flux $\dot{M}_{\rm r}$ (bottom panel) of dust particles 
as a function of orbital radius $r$ for models with $r_{\rm out}  = 100~{\rm AU}$ 
and with different values of $\alpha$.
The dashed and dotted lines are the snapshots at times $t =$ 0.1 and 5 Myr, respectively.
}
\label{fig:daalpha}
\end{figure}

The top and middle panels of Figure~\ref{fig:dr} {show} the global evolution 
of the dust surface density $\Sigma_{\rm d}$
and particle size $a$ for $\alpha = 10^{-3}$.
The left and right panels correspond to small and large disks
with $r_{\rm out} =$ 100 AU and 300 AU, respectively.
The radial mass flux of the particles, $\dot{M}_{\rm d}$ (Eq.~\eqref{eq:Mdotr}), 
is shown in the bottom panels.   
The results for different values of $\alpha$ are presented in Fig.~\ref{fig:daalpha}.
One can see that dust particles grow significantly and are piled up at $r \approx 1~{\rm AU}$
in the very early stage of $t = 0.1~{\rm Myr}$.
This occurs because dust particles initially located at these inner orbits 
grow beyond the radial drift barrier owing to accelerated coagulation in the Stokes regime \citep{BDB10,OTKW12}. 
However, this feature is immediately erased 
by the significant amount of pebbles flowing from outer disk regions. 
{Furthermore, in this very early stage, the snow line would be in reality well outside 1 AU.
If this is the case, dust particles near 1 AU would be made of silicates rather than water ice,
and their collisional growth would be limited by fragmentation \citep{BW08,WTS+09}.
}  
Since we assume that pebble accretion starts only after $t =0.5~{\rm Myr}$,  
this early feature has no effect on the results of our pebble accretion calculations.

As demonstrated by many previous studies \citep[e.g.,][]{TL05,G07,BDH08,BDB10,BKE12,OTKW12},
global dust evolution can be understood from timescale arguments 
as presented in Sect.~\ref{sec:velocity}.
In protoplanetary disks, dust growth commences from inside out 
because the growth timescale $t_{\rm grow}$ (Eq.~\eqref{eq:tgrow}) 
is roughly proportional to the orbital period.
At each orbital distance, dust particles initially grow without appreciable drift, 
conserving the dust surface density at that location. 
This local growth stage continues until the particles acquire a high drift velocity.
Once the drift timescale becomes comparable to the growth timescale, 
the particles start drifting inward so that the two timescales balance each other.  
In this second stage, the dust surface density at each location is no longer locally conserved
and is instead determined by the mass flow of particles drifting from further out. 
To an order of magnitude, the time required for initially micron-sized particles to grow 
into drifting pebbles is estimated as $\sim 10t_{\rm grow}$, where the factor $10$ 
accounts for the fact that the particles need to grow by several orders of 
magnitude in size to acquire a high drift velocity \citep{LJ14}.  
If we take $t_{\rm grow} \sim 100/\Omega$ (see Eq.~\eqref{eq:tgrow_approx}),
we have $10t_{\rm grow} \sim 0.1$ Myr at $r=$ 60 AU and $10t_{\rm grow} \sim 1$ Myr at $r =$ 300 AU.
This is consistent with the results shown in Figure~\ref{fig:dr}, 
where we can see that the radial dust flow originates at $\approx 60~{\rm AU}$ 
and $\approx 300~{\rm AU}$ for $t = 0.1~{\rm Myr}$ and 1 Myr, respectively. 
However, the growth timescale also depends on turbulence strength $\alpha$
as already noted in Sect.~\ref{sec:velocity}. 
For example, we can see in Fig.~\ref{fig:daalpha} that dust particles at 100 AU have already grown 
significantly even at 0.1 Myr in the case of $\alpha = 10^{-2}$.
This is because of the short growth timescale at $a \la 10^{-2}~{\rm cm}$ 
for this value of $\alpha$ (see Fig.~\ref{fig:tgrow}).

A key parameter that controls the global dust evolution is the radial extent of the initial dust disk, $r_{\rm out}$. 
In a typical protoplanetary disk with a surface density gradient $d\ln\Sigma_{\rm g}/d\ln r > -2$, 
the dominant part of the disk mass resides in outer regions of the disk.
The outer edge of a disk thus generally acts as a dust reservoir that produces inwardly drifting pebbles \citep{G07,BKE12,LJ14}.
For example, one can see in the top panels of Fig.~\ref{fig:dr} that the dust surface density 
$\Sigma_{\rm d}$ starts decreasing at all orbital distances as the outer edge of the disk gets depleted of dust. 
The pebble size $a$ decreases at the same time, since the growth timescale becomes longer and longer 
as $\Sigma_{\rm d}$ declines (see Eq.~\eqref{eq:tgrow}). 
The lifetime of this dust reservoir is essentially determined by the growth timescale 
of the dust at $r \sim r_{\rm out}$, and hence increases with $r_{\rm out}$.
This explains why the pebble flow in the $r_{\rm out} = 100~{\rm AU}$ disk diminishes 
faster than in the $r_{\rm out} = 300~{\rm AU}$ disk.
Our numerical simulations show that dust depletion starts at $t \approx 0.2~{\rm Myr}$
for $r_{\rm out} = 100~{\rm AU}$ and at $t \approx 1~{\rm Myr}$
for $r_{\rm out} = 300~{\rm AU}$.

\subsection{Size and mass flux of pebbles at 1~AU}\label{sec:1AU}
In the context of pebble accretion, the quantities of interest are the size and radial mass flux 
of drifting particles at the embryo's orbit.  
Figure~\ref{fig:aM} shows these quantities at the Earth's orbit, $r = 1~{\rm AU}$. 
As explained in the previous subsection, the size and surface density of the particles 
decrease as the outer region of the disk is depleted of dust.  
Before this depletion occurs, particles arriving at 1 AU have a nearly constant radius 
$a \approx 20$--30 cm and a nearly constant Stokes number ${\rm St} \approx 0.2$--$0.5$,
which is consistent with the results of previous studies \citep{BDH08,BKE12,OTKW12,LJ14}.
The radial mass flux at this early time is on the order of $10^{2}~M_\oplus~{\rm Myr}^{-1}$.
This directly follows from fact that the dust in outer disk regions 
has a total mass of $\sim 10^2 M_\oplus$ (see Eq.~\eqref{eq:Md}) and grows into drifting pebbles 
on a timescale of $\sim 10~t_{\rm grow}|_{r=r_{\rm out}}  \sim 1~{\rm Myr}$ (see Sect.~\ref{sec:global}).

Once the dust depletion at the outer edge begins, the particle size and radial flux decrease with time. 
The decrease of the particle size can be understood from the competition 
between coagulation and radial drift.
In general, dust particles are allowed to grow as along as the 
growth timescale is shorter than the drift timescale $\sim r/|v_{\rm r}| \propto |v_{\rm r}|^{-1}$.  
As $\Sigma_{\rm d}$ decreases, the growth timescale increases 
(since $t_{\rm grow} \propto \Sigma_{\rm d}^{-1}$), 
and consequently balances with the drift timescale at smaller particle size  
(since $|v_{\rm r}| \propto {\rm St} \propto a$).  

\subsection{Pebble accretion rate}\label{sec:Medot}

Now we apply the results presented in Sect.~\ref{sec:1AU} to pebble accretion by an embryo 
located at the Earth's orbit. To begin with, we calculate the pebble accretion rate $\dot{M}_{\rm e}$
(Eq.~\eqref{eq:Medot}) of an embryo of {\it fixed} mass ${M}_{\rm e}$. 
The evolution of ${M}_{\rm e}$ is discussed in Sect.~\ref{sec:evol}.

\begin{figure*}[t]
\centering
\includegraphics[width=17cm]{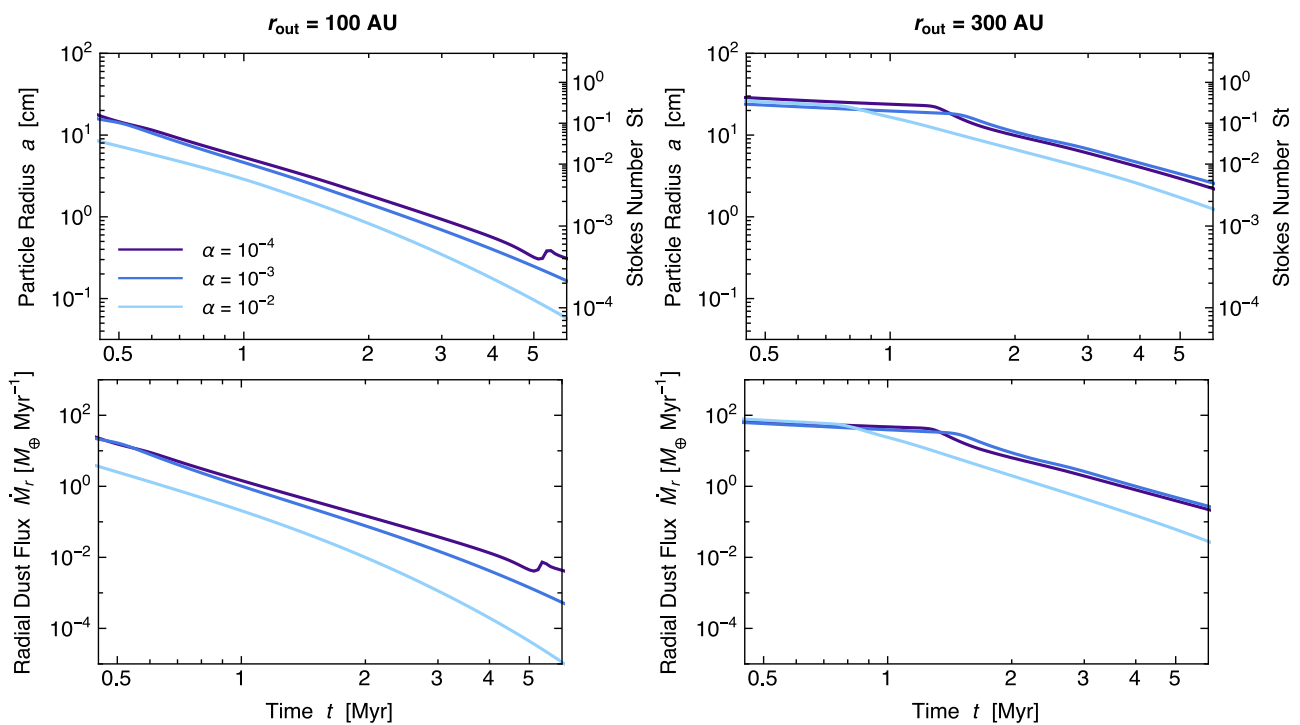}
\caption{Radius $a$ (upper panels) and radial mass flux $\dot{M}_{\rm r}$ (lower panels) 
of drifting particles observed at 1~AU as a function of time $t$. 
The left and right panels are for $r_{\rm out} =$ 100 AU and 300 AU, respectively.
}
\label{fig:aM}
\end{figure*}

\begin{figure*}[t]
\centering
\includegraphics[width=17cm]{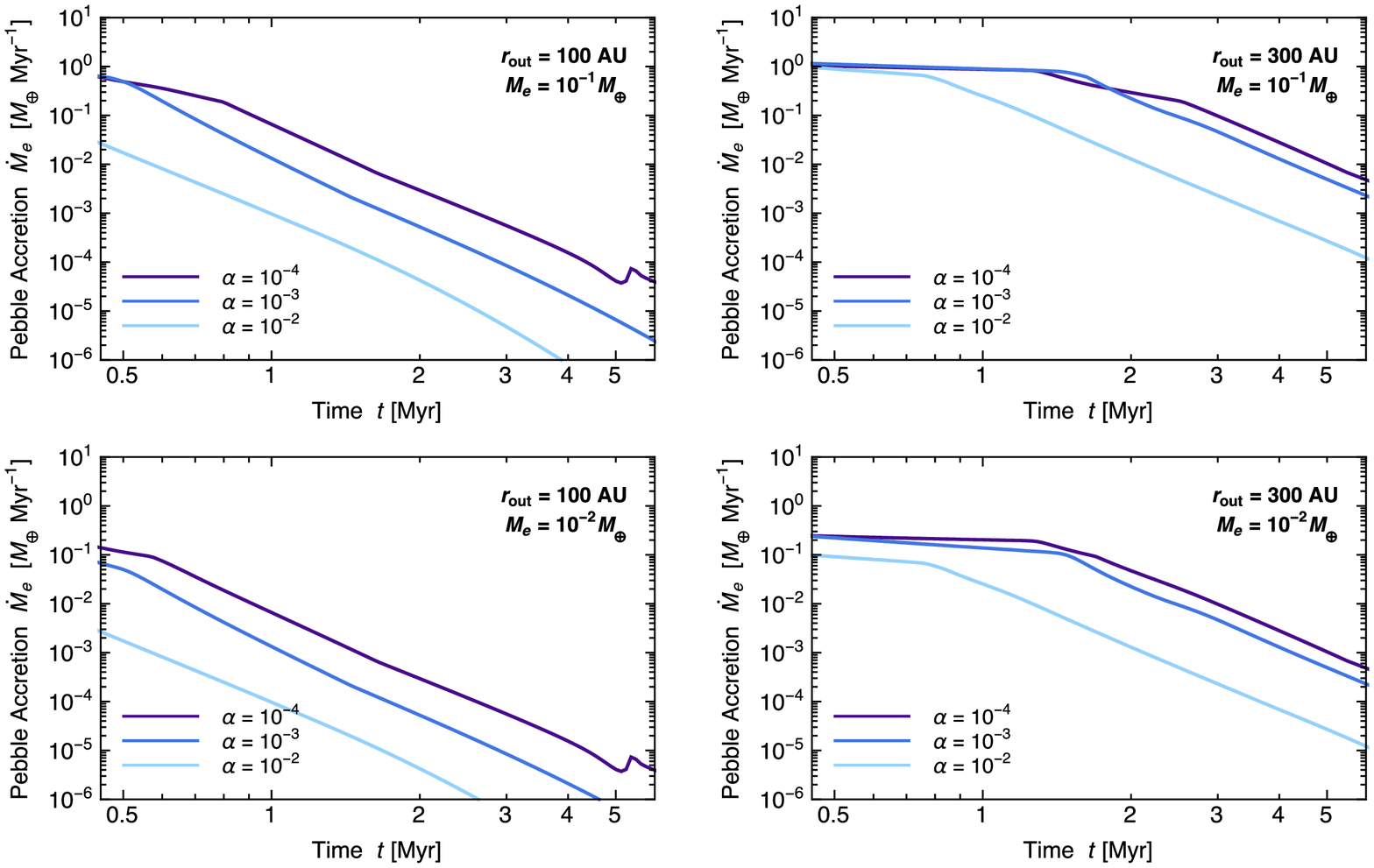}
\caption{Pebble accretion rate $\dot{M}_{\rm e}$ of a single embryo (Eq.~\eqref{eq:Medot})
located at 1 AU as a function of time $t$ for $M_{\rm e} = 10^{-1}~M_\oplus$ (upper panels)
and $10^{-2}~M_\oplus$ (lower panels). 
The left and right panels are for $r_{\rm out} =$ 100 AU and 300 AU, respectively.
}
\label{fig:Medot}
\end{figure*}

In the upper panels of Figure~\ref{fig:Medot}, we plot $\dot{M}_{\rm e}$ of an embryo with 
$M_{\rm e} = 10^{-1}M_\oplus$ as a function of time $t$ for different values of $r_{\rm out}$.
Again, the result significantly depends on the value of $r_{\rm out}$ as it determines the lifetime 
of the dust reservoir at the disk outer edge. 
At early times when a substantial amount of dust remains at the outer edge ($t \la 0.5$ Myr for $r_{\rm out} = 100~{\rm AU}$ and $t \la 1$ Myr for $r_{\rm out} = 300~{\rm AU}$), particles drifting to 1 AU are 20--30 cm in size (see the upper panels of Fig.~\ref{fig:aM}) and, hence, are swept up by a single $10^{-1}M_\oplus$ embryo at a probability of $\approx$ 1--2\% (Fig.~\ref{fig:P}). 
Since the radial mass flux of these decimeter-sized particles is $10^2M_\oplus~{\rm Myr}^{-1}$ 
(the lower panels of Fig.~\ref{fig:aM}), the accretion probability of about 1\%  
results in an accretion rate of $\approx 1 M_\oplus~{\rm Myr}^{-1}$ as shown in Fig.~\ref{fig:Medot}.
This value is insensitive to the choice of $\alpha$, as long as $\alpha \leq 10^{-2}$, 
because the particle accretion is nearly two-dimensional ($h_{\rm d} \la b_{\rm set}$) 
at these particle sizes (see Fig.~\ref{fig:P}).
As the dust in the outer disk is depleted, $\dot{M}_{\rm e}$ decreases with decreasing $\dot{M}_{\rm r}$.
In this late stage, $\dot{M}_{\rm e}$ becomes more sensitive to $\alpha$
with a higher $\alpha$ resulting in an even smaller $\dot{M}_{\rm e}$.
This is mainly because the smaller drifting particles in this stage 
accrete onto the embryo in a 3D manner (i.e., $h_{\rm d} > b_{\rm set}$), 
for which case a higher $\alpha$ results in a lower  $\dot{M}_{\rm e}$.
We can see that $\dot{M}_{\rm e}$ for $\alpha = 10^{-2}$ 
is approximately 10--100 times smaller than that for $\alpha = 10^{-4}$ in this stage.

The results for ${M}_{\rm e} = 10^{-2} M_\oplus$ (the lower panels of Fig.~\ref{fig:Medot})
are qualitatively similar to those for ${M}_{\rm e} = 10^{-1} M_\oplus$
except that the magnitude of $\dot{M}_{\rm e}$ is decreased by a factor of 5--10.
As we show below, 
this directly follows  from the dependence of the pebble accretion radius $b_{\rm set}$ on $M_{\rm e}$. 
Comparison between Eq.~\eqref{eq:St_bnd} and the Stokes number plotted in Fig.~\ref{fig:aM} 
shows that pebble accretion occurs in the Hill accretion regime for both values of $M_{\rm e}$.
Since $b_{\rm set} \propto R_{\rm H} \propto M_{\rm e}^{1/3}$ and 
$v_{\rm set} \propto b_{\rm set} \propto M_{\rm e}^{1/3}$ in this accretion regime (see Sect.~\ref{sec:pebble}), 
we obtain $\dot{M}_{\rm e} \propto M_{\rm e}^{2/3}$ in the 2D case ($b_{\rm set} \la h_{\rm d}$) 
and $\dot{M}_{\rm e} \propto M_{\rm e}$ in the 3D case  ($b_{\rm set} \ga h_{\rm d}$).
Therefore, decreasing $M_{\rm e}$ by the factor of 10 results in a decrease in 
$\dot{M_{\rm e}}$ by a factor of $10^{2/3}$--10 $\approx 5$--10.

It is worth mentioning at this point that the timescale of embryo growth by pebble accretion, 
$M_{\rm e}/\dot{M}_{\rm e}$, is a weak function of the embryo mass: 
$M_{\rm e}/\dot{M}_{\rm e} \propto M_{\rm e}^{1/3}$ in the 2D case and 
$M_{\rm e}/\dot{M}_{\rm e}\propto M_{\rm e}^{0}$ in the 3D case.
This implies that the rate at which the embryo's water mass fraction
increases is insensitive to the choice of $M_{\rm e}$. 
We confirm this expectation in the following subsection. 

\subsection{Evolution of embryo's mass and water fraction}\label{sec:evol}
We now let an embryo grow through icy pebble accretion to 
study how much water is delivered to the embryo from icy pebbles. 
We place a rocky embryo initially devoid of water at 1 AU
and allow it to start accreting icy pebbles at  $t = t_{\rm start}$.   
We calculate the evolution of the embryo mass $M_{\rm e}$ at $t > t_{\rm start}$
by integrating Eq.~\eqref{eq:Medot} taking 
the change in the accretion radius $b_{\rm set}$ with the change in  $M_{\rm e}$ into account. 
The evolution of the embryo's water fraction, $f_{\rm H_2O}$, 
is computed assuming that the water content of the accreted 
pebbles is 50~wt\%, i.e.,
\beq
f_{\rm H_2O}(t) \equiv \frac{ \int_{t_{\rm start}}^{t} 0.5 \dot{M}_{\rm e}(t') dt' }
{M_{\rm e,0} +\int_{t_{\rm start}}^{t}\dot{M}_{\rm e}(t') dt' } 
= \frac{M_{\rm e}(t) - M_{\rm e,0}}{2M_{\rm e}(t)},
\label{eq:f}
\eeq
where $M_{\rm e,0}$ is the initial embryo mass and 
$M_{\rm e}(t)$ is the embryo mass at time $t$ $(> t_{\rm start})$. 
We have assumed that $f_{\rm H_2O} = 0$ in the initial state.

Table~\ref{tb:results} lists the mass and water content in the final state (taken to be $t = 6~{\rm Myr}$) 
for various sets of model parameters (see Table~\ref{tb:parameters} for the parameter grid). 
We immediately find that the final water fraction is insensitive to $M_{\rm e,0}$,
which is because the scaled pebble accretion rate $\dot{M}_{\rm e}/M_{\rm e}$
is nearly independent of $M_{\rm e}$ as already noted in Sect.~\ref{sec:Medot}. 
In the following, we focus on the results for $M_{\rm e} = 10^{-1}M_\oplus$. 

\begin{table}
\centering
\begin{tabular}{llllll}
\hline\hline
$r_{\rm out}$ & $\alpha$ & $M_{\rm e,0}$  & $t_{\rm start}$ &
$M_{\rm e, 6Myr}$  & $f_{\rm H_2O, 6Myr}$  \\
(AU) &  & ($M_{\oplus}$) & (Myr) & ($M_{\oplus}$) & (wt\%) \\  \hline
100 & $10^{-4}$ & 0.1 & 0.5 & 0.39 & 37 \\ 
 & &  & 1 & 0.13 & 12 \\ 
 & &  & 2 & 0.10 & 1.5 \\ 
 & &  & 4 & 0.10 & 0.14 \\ \hline
100 & $10^{-4}$ & 0.01 & 0.5 & 0.11 & 45 \\ 
 & &  & 1 & 0.014 & 15 \\ 
 & &  & 2 & 0.010 & 2.2 \\ 
 & &  & 4 & 0.010 & 0.21 \\ \hline
100 & $10^{-3}$ & 0.1 & 0.5 & 0.20 & 26 \\ 
 & &  & 1 & 0.11 & 2.5 \\ 
 & &  & 2 & 0.10 & 0.25 \\ 
 & &  & 4 & 0.10 & 0.017 \\ \hline
100 & $10^{-3}$ & 0.01 & 0.5 & 0.024 & 29 \\ 
 & &  & 1 & 0.011 & 3.4 \\ 
 & &  & 2 & 0.010 & 0.38 \\ 
 & &  & 4 & 0.010 & 0.028 \\ \hline
100 & $10^{-2}$ & 0.1 & 0.5 & 0.10 & 1.8 \\ 
 & &  & 1 & 0.10 & 0.23 \\ 
 & &  & 2 & 0.10 & 0.018 \\ 
 & &  & 4 & 0.10 & 0.00063 \\ \hline
100 & $10^{-2}$ & 0.01 & 0.5 & 0.010 & 2.3 \\ 
 & &  & 1 & 0.010 & 0.32 \\ 
 & &  & 2 & 0.010 & 0.028 \\ 
 & &  & 4 & 0.010 & 0.0011 \\ \hline
300 & $10^{-4}$ & 0.1 & 0.5 & 19. & 50 \\ 
 & &  & 1 & 7.8 & 49 \\ 
 & &  & 2 & 1.2 & 46 \\ 
 & &  & 4 & 0.14 & 15 \\ \hline
300 & $10^{-4}$ & 0.01 & 0.5 & 15. & 50 \\ 
 & &  & 1 & 5.5 & 50 \\ 
 & &  & 2 & 0.52 & 49 \\ 
 & &  & 4 & 0.016 & 19 \\ \hline
300 & $10^{-3}$ & 0.1 & 0.5 & 25. & 50 \\ 
 & &  & 1 & 10. & 49 \\ 
 & &  & 2 & 0.54 & 41 \\ 
 & &  & 4 & 0.12 & 7.5 \\ \hline
300 & $10^{-3}$ & 0.01 & 0.5 & 20. & 50 \\ 
 & &  & 1 & 7.0 & 50 \\ 
 & &  & 2 & 0.073 & 43 \\ 
 & &  & 4 & 0.012 & 9.7 \\ \hline
300 & $10^{-2}$ & 0.1 & 0.5 & 4.2 & 49 \\ 
 & &  & 1 & 0.25 & 30 \\ 
 & &  & 2 & 0.11 & 5.0 \\ 
 & &  & 4 & 0.10 & 0.49 \\ \hline
300 & $10^{-2}$ & 0.01 & 0.5 & 0.76 & 49 \\ 
 & &  & 1 & 0.029 & 33 \\ 
 & &  & 2 & 0.011 & 6.4 \\ 
 & &  & 4 & 0.010 & 0.70 \\ \hline
\end{tabular}
\caption{Mass $M_{\rm e}$ and water content $f_{\rm H_2O}$ of a rocky embryo 
at 1AU for various sets of model parameters. The values are measured at time $t = 6~{\rm Myr}$
after disk formation.}
\label{tb:results}
\end{table}

\begin{figure*}[t]
\centering
\includegraphics[width=17cm]{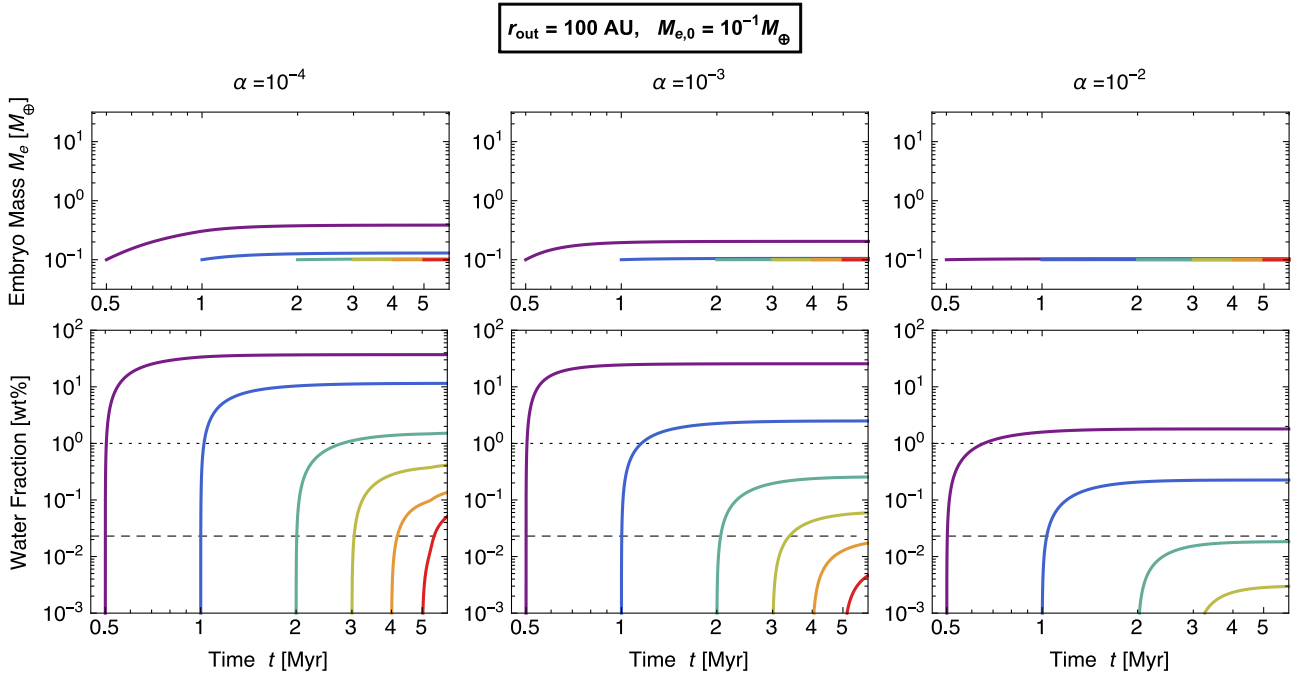}
\caption{Time evolution of the mass $M_{\rm e}$ and water fraction $f_{\rm H_2O}$ 
of an embryo placed at 1 AU with initial mass $M_{\rm e,0} = 10^{-1}M_\oplus$ 
for the case of $r_{\rm out} = 100~{\rm AU}$. 
The different curves show results for different initial times of pebble accretion, 
$t_{\rm start} =$ 0.5, 1, 2, 3, 4, and 5 Myr (from left to right).
}
\label{fig:wf11}
\end{figure*}
\begin{figure*}[t]
\centering
\includegraphics[width=17cm]{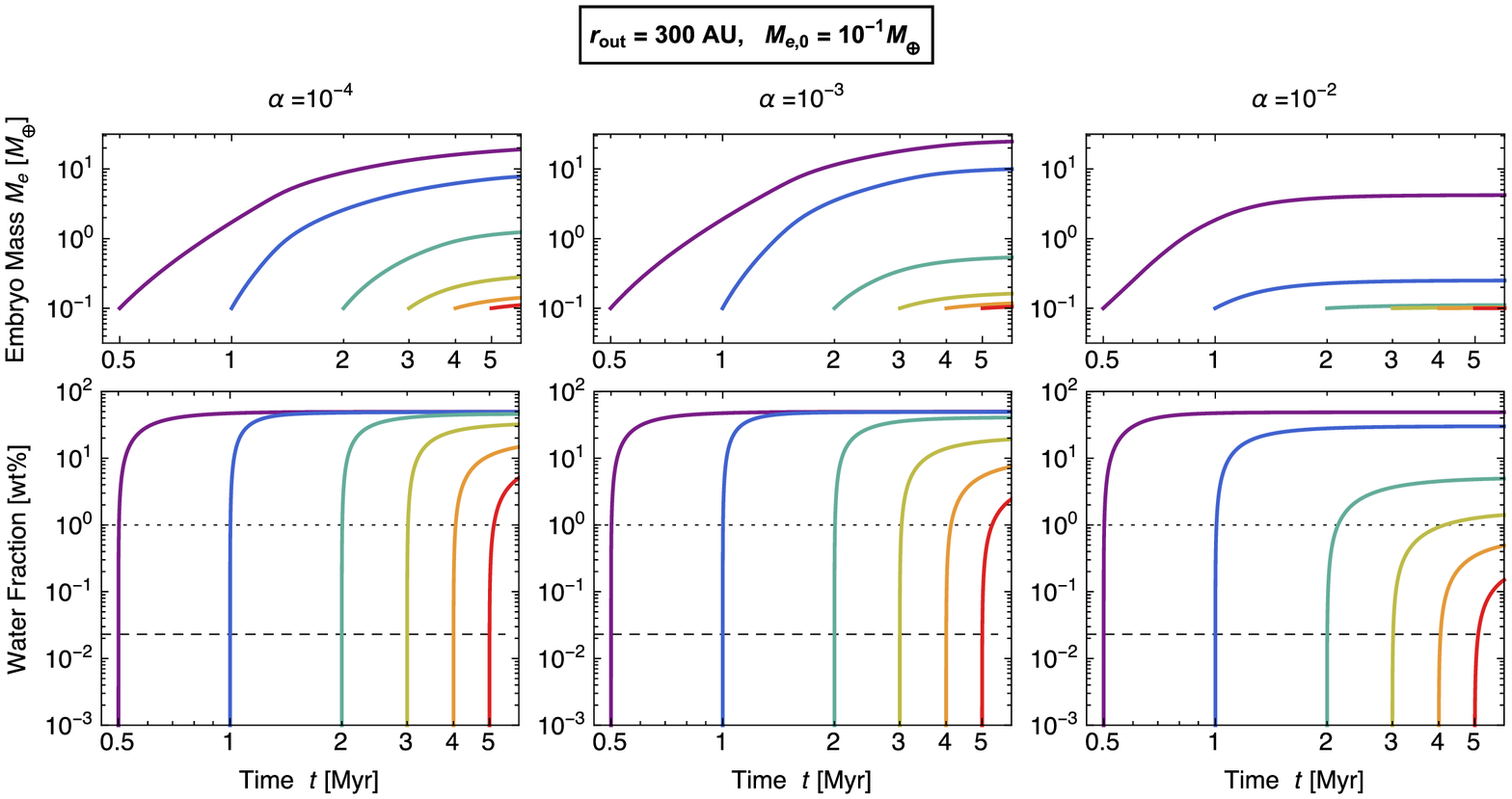}
\caption{Same as Figure~\ref{fig:wf11}, except for $r_{\rm out} = 300~{\rm AU}$.}
\label{fig:wf12}
\end{figure*}

Figure~\ref{fig:wf11} shows the evolution of $M_{\rm e}$ and $f_{\rm H_2O}$ 
in the $r_{\rm rout} = 100~{\rm AU}$ disk model for various values of $\alpha$ and $t_{\rm strart}$.
The results for the larger disk model ($r_{\rm rout} = 300~{\rm AU}$) are shown in Fig.~\ref{fig:wf12}. 
In the plots of $f_{\rm H2O}$, the dashed lines indicate the minimum water content of the present Earth 
given by the ocean mass, 0.023 wt\%. The dotted lines indicate $f_{\rm H_2O} = 1~{\rm wt\%}$,
which corresponds to the hypothetical water content of the proto-Earth assuming that 
the density deficit of the outer core is due to hydrogen delivered in the form of water \citep{O97,AOORD00}. 
The water content of the proto-Earth much in excess of $\sim 1\%$ seems unlikely \citep{MA10}.

In the case of $r_{\rm rout} = 100~{\rm AU}$, whether Earth-forming rocky embryos 
avoid excessive ice accretion depends on the values of $\alpha$ and $t_{\rm start}$. 
For $\alpha = 10^{-4}$ (the left panels of Fig.~\ref{fig:wf11}), 
the embryo's final water content exceeds 0.023 wt\% for all $t_{\rm start} \la 5~{\rm Myr}$. 
It even exceeds 1 wt\% if $t_{\rm start}$ is as short as $\la 2~{\rm Myr}$.
In the extreme case of $t_{\rm start}=0.5~{\rm Myr}$, the final embryo mass is  
four times larger than the initial mass, meaning that the rocky embryo has evolved 
into an icy embryo mostly composed of icy pebbles. 
The embryo acquires a smaller amount of water if the disk is more turbulent ($\alpha$ is higher).
For $\alpha = 10^{-3}$ and $10^{-2}$ (the middle and right panels of Fig.~\ref{fig:wf11}), 
the final $f_{\rm H_2O}$ does not exceed 1\% if $t_{\rm start} > 1~{\rm Myr}$ and 0.5~Myr, respectively. 
A water fraction of $\la$ 0.023 wt\% is achieved if $t_{\rm start} > 4~{\rm Myr}$ for $\alpha=10^{-3}$
and if $t_{\rm start}$ > 2~Myr for $\alpha = 10^{-2}$.
This reduction is due to the combination of the accelerated coagulation and 3D pebble accretion 
already discussed in Sect.~\ref{sec:Medot}.

Preserving a rocky embryo from icy pebbles is much more difficult when the gas disk extends to 
$300~{\rm AU}$ (Fig.~\ref{fig:wf12}).
In this case, no parameter set is found to result in a final water content smaller than 0.023 wt\%.
Even a final water content of $\la 1$ wt\% is realized only if $\alpha = 10^{-2}$ and $t_{\rm start} \ga 3~{\rm Myr}$.  
Instead, we find that the initially rocky embryo evolves into a super-Earth to Neptune-mass icy planet 
if  $t_{\rm start} \la 0.5$--$2~{\rm  Myr}$ (the smaller and large values correspond to $\alpha = 10^{-2}$ and $10^{-4}$, respectively).
Our results for $r_{\rm out} = 300~{\rm AU}$ and $t_{\rm start} = 0.5~{\rm Myr}$ 
are similar to the results of \citet{LJ14} for giant planet core formation in outer disk regions. 
This is reasonable because the pebble flow of \citet{LJ14} is assumed to decay on the timescale of $3~{\rm Myr}$, while the pebble flow in our $r_{\rm out} = 300~{\rm AU}$ calculations decays on a similar timescale. 

In summary, we find that embryos orbiting at 1 AU can remain rocky at a level of
$f_{\rm H_2O} \la 0.023~{\rm wt\%}$ if the disk size is 100 AU or smaller, 
turbulence is stronger than $\alpha = 10^{-3}$, and the snow line passes 1 AU 
later than 2--4 Myr after disk formation. 
Keeping the water fraction at a level of $f_{\rm H_2O} \la 1~{\rm wt\%}$ with a disk size of 100 AU
is possible if the snow line migrates in after $t=0.5$--2 Myr.   
If the disk is as large as 300 AU, a final water fraction of $\la 0.024~{\rm wt\%}$ is very unlikely, 
and a final fraction of $\la 1~{\rm wt\%}$ is possible 
only if turbulence is strong ($\alpha = 10^{-2}$) and if the snow line migrates later than $3~{\rm Myr}$.
 
\section{Discussion}\label{sec:discussion}

\begin{figure}[t]
\centering
\includegraphics[width=9cm]{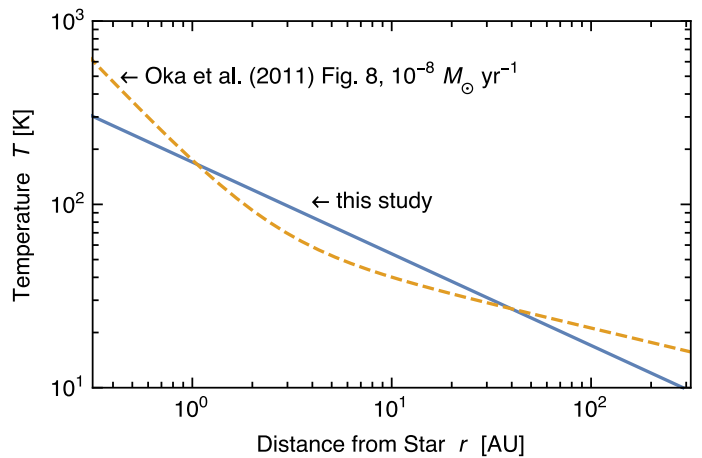}
\caption{Temperature profiles adopted in this study (Eq.~\eqref{eq:T}; solid line)
and from a viscous accretion model of \citet{ONI11} (Eq.~\eqref{eq:T_ONI11}; dashed curve).}
\label{fig:T}
\end{figure}

\begin{figure*}[t]
\centering
\includegraphics[width=14cm]{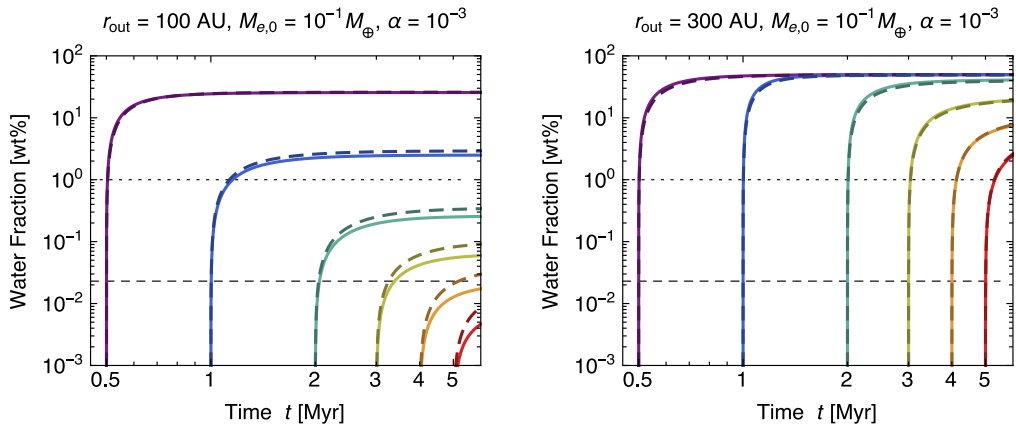}
\caption{Same as the lower center panels of Figs.~\ref{fig:wf11} and \ref{fig:wf12},
but here the results for the temperature profile from an optically thick viscous disk model
(Eq.~\eqref{eq:T_ONI11}) are overplotted (dashed curves). 
}
\label{fig:wf_ONI11}
\end{figure*}

{
\subsection{Dependence on the temperature profile}\label{sec:T}
We have simplified the radial temperature profile $T(r)$ 
with a single power law for an optically thin disk (Eq.~\eqref{eq:T}).
In an optically thick disk, the temperature profile is steeper in inner regions where viscous heating dominates
and is shallower in outer regions where stellar irradiation dominates. 
Detailed modeling of the temperature profile is beyond the scope of this paper,
but we show below that our results are fairly insensitive to the choice of the temperature profile.

We adopt the temperature profile of an optically thick disk around 
a Sun-like star presented by \citet{ONI11}. 
We select one of their models in which $\dot{M} = 10^{-8} M_\odot~{\rm yr}^{-1}$
and $\alpha = 10^{-3}$ with a dust opacity mimicking that of \citet{GL07}.   
We chose this model because the midplane temperature reaches 170 K at 1 AU 
as in our fiducial temperature profile. 
The radial profile of the midplane temperature for this model is shown in their Fig.~8 (black solid line). 
We find that this profile can be reasonably reproduced by a simple analytic fit 
\beq
T(r) = \sqrt{[160 (r/1~{\rm AU})^{-1.15}]^2 + [70 (r/1~{\rm AU})^{-0.26}])^2}~{\rm K},
\label{eq:T_ONI11}
\eeq
which is shown by the dashed curve in our Fig.~\ref{fig:T}. 
In this model, viscous heating is effective at $\la$ 4 AU and the temperature in that region 
rises toward the central star as steeply as $T \propto r^{-1.15}$.
However, as far as the region $1~\AU \leq r \leq 300~\AU$ is concerned,   
the difference in the values of $T$ between the two models is small
with the maximum deviation of $\approx$ 60 \%. 
The surface density profile in this viscous disk model differs from 
the MMSN model (Eq.~\eqref{eq:Sigmag}). 
However, we keep using the MMSN density profile to isolate 
the effects of changing the temperature profile.

We fix $M_{\rm e,0} = 10^{-1}M_\oplus$ and $\alpha = 10^{-3}$
and only vary the values of $r_{\rm out}$ and $t_{\rm start}$. 
The results for the two different temperature models are compared in Fig.~\ref{fig:wf_ONI11}. 
Here, the solid curves show the evolution of the embryo water fraction $f_{\rm H_2O}$ 
for our temperature profile  (Eq.~\eqref{eq:T}), which is already shown in the lower center panels 
of Figs.~\ref{fig:wf11} and \ref{fig:wf12}, while 
the dashed curves are for the viscous disk temperature profile given by Eq.~\eqref{eq:T_ONI11}.
One can see that the predicted water fraction of the embryo 
is very insensitive to the choice of the temperature profile.
A closer inspection shows that in the $r_{\rm out} = 100~\AU$ case (the left panel
of Fig.~\ref{fig:T}), the viscous temperature model leads to a slightly higher $f_{\rm H_2O}$,
in particular, at $t \ga 2~\rm Myr$ at which the radial pebble flux has already decayed to 
$0.1 M_\oplus~{\rm Myr^{-1}}$ (see the lower left panel of Fig.~\ref{fig:aM}).
However, the final values of $f_{\rm H_2O}$ in the viscous model 
are only larger than those in our fiducial model by a factor of less than 2.
We conclude that the details of the assumed temperature profile do not affect our conclusions.  
}

\subsection{Migration timescale of the snow line}\label{sec:gasst}
We have shown in Sect.~\ref{sec:results} that the fate of terrestrial embryos 
largely depends on the timing of inward snow-line migration, $t_{\rm start}$.
Rocky embryos are able to avoid excessive icy pebble accretion only 
if the snow line migrates in {\it after} the radial pebble flux in the disk is sufficiently depleted. 
This is already obvious from the estimate of the pebble accretion rate 
presented in Sect.~\ref{sec:Medot}.
The pebble accretion rate of a 0.1 Earth mass embryo before dust depletion 
is $\sim 1~M_\oplus~{\rm Myr}^{-1}$, which 
roughly amounts to 0.1 Earth ocean mass ($\sim 10^{-5}M_\oplus$) in 10 years.
If the Earth-forming embryos contained less water than on the ocean of the present Earth 
(water content $\la 0.023~{\rm wt\%}$), the snow line must have migrated to 1 AU
as late as 2--4 Myr after nebula formation 
(assuming that the nebula had a radial extent of $\sim 100~{\rm AU}$; see  Fig.~\ref{fig:wf11}). 
Even if the Earth formed from wetter embryos of water content $1~{\rm wt\%}$ \citep{MA10}, 
the migration of the snow line must have occurred no earlier than 0.5--2 Myr.

The remaining question is then whether these conditions can be
satisfied in a realistic protoplanetary disk.
Addressing this questing with a detailed model of snow-line migration is beyond the scope of this paper.
Here we attempt to estimate the timescale of snow-line migration 
assuming that (i) viscous heating dominates over stellar irradiation, 
and that (ii) the disk opacity is constant in time. 
In this simplest case, the timescale of snow-line evolution is essentially given by 
the viscous evolution timescale of the disk,
\begin{equation}
t_{\rm visc} \sim \frac{r_{\rm out}^2}{\nu|_{r=r_{\rm out}}} \sim 2~{\rm Myr} 
\left(\frac{\alpha|_{r=r_{\rm out}}}{10^{-2}}\right)^{-1}
\left(\frac{r_{\rm out}}{100~\AU}\right),
\label{eq:tvisc}
\end{equation}
where $\nu = \alpha c_{\rm s}h_{\rm g}$ is the turbulent viscosity
and we have used Eq.~\eqref{eq:T} in the final expression. 
If we take $r_{\rm out} = 100~\AU$ and $\alpha|_{r=r_{\rm out}} = 10^{-2}$, 
we obtain $t_{\rm visc} \sim 2~{\rm Myr}$,
which is comparable to the time required for sufficient dust depletion.
Therefore, snow-line migration after the decay of the radial pebble flow 
is a possible explanation for the origin of the dry Earth. 
However, Eq.~\eqref{eq:tvisc} only serves as a rough estimate of $t_{\rm start}$, 
and a more precise assessment taking into account viscous evolution, stellar evolution, 
and the evolution of the disk opacity due to dust evolution is necessary.
This will be addressed in future work.

We point out that $t_{\rm visc} \propto r_{\rm out}$ whereas 
$t_{\rm grow}|_{r=r_{\rm out}} \propto \Omega^{-1}|_{r=r_{\rm out}} \propto r_{\rm out}^{3/2}$. 
This implies that when $r_{\rm out}$ is small, 
the snow line tends to migrate more slowly than icy dust in the disk becomes depleted 
(whose timescale is $\propto t_{\rm grow}$ at $r=r_{\rm out}$).
This argument also supports the idea that preservation of rocky embryos from ice pebbles 
favors a compact protoplanetary disk.

{On the other hand, an extended disk is beneficial for forming the cores of gas giants 
at wider orbits through the pebble accretion mechanism \citep{LJ14}.
As we discuss below, such fully grown cores could save the rocky embryos in the inner disk
by halting the ice pebble flow.
}

\subsection{Possible mechanisms for pebble filtration outside 1 AU} \label{sec:filter}
We have restricted ourselves to the simplest (and most pessimistic) case 
where all icy pebbles forming in outer disk regions are allowed to drift to 1 AU. 
In fact, there are some known mechanisms that might halt or filter the pebble flux before 
they reach rocky embryos. Ignoring such possibilities effectively means that we have assumed 
these mechanisms operate only after the snow line migrates to 1 AU. 
We here mention some import mechanisms and discuss whether they are likely to have operated in the solar nebula.

The most straightforward scenario would be that
planetesimals or embryos outside the Earth's orbit filter out incoming pebbles
just in the same way as what we considered for Earth-forming embryos.
However, it turns out an efficient filtration is not expected with this mechanism in many cases.
We have shown in Sect.~\ref{sec:Medot} that each $0.1M_\oplus$-mass embryo captures
only $\approx 1$--$2~\%$ of the pebble flux in a disk. 
Assuming that the total mass of terrestrial embryos in the minimum-mass solar nebula 
is $\approx 2~M_\oplus$ (Earth + Venus), 
the total number of the embryos is $\approx 20$, and therefore
they filter only $\approx 20$--$40~\%$ of the pebble flux in total. 
The low ($< 50\%$) efficiency of dust filtration by a small number 
of embryos {are consistent with the results by} \citet{MN12}, \citet{LJ14},  {and \citet{MLJB15}}.
A more systematic study on pebble filtration by planetesimal- or embryo-sized objects
has been carried out by \citet{GIO14} using essentially the same pebble accretion formula as ours. 
They found that perfect filtration beyond 1 AU is possible 
only if most of the dust in the planet-forming region 
is converted into $\sim 1000~{\rm km}$-sized embryos
and if disk turbulence is $\alpha = 10^{-4}$ (see their Fig.~22).  
{\cite{MF15} obtained qualitatively similar results; these authors 
considered the accretion of cm-sized drifting pebbles by initially 100 km-sized planetesimals
in an $\alpha = 10^{-3}$ gas disk and showed that the inward flux of cm-sized pebbles 
is nearly constant down to $1~{\rm AU}$ (see their Fig.~2).}
These suggest that a swarm of outer planetesimals and embryos is only 
able to fully filter the icy pebble flow in a particular range of parameter space. 
Whether such a situation was realized in the solar nebula 
over the lifetime of pebble flow ($\sim 0.5$--$2~{\rm Myr}$) is unclear.

A more promising mechanism for pebble filtration
is particle trapping at pressure maxima in the gas disk.
In general, the direction of particle radial drift is determined by the sign of 
the pressure gradient of the disk (Eq.~\eqref{eq:eta}), 
and therefore particles tend to accumulate toward locations 
where the gas pressure is locally maximized \citep{W72}. 
A pressure bump may be created by magnetorotational turbulence \citep[e.g.,][]{JYK09,UKFH11}, 
by a steep gradient  in the gas viscosity \citep[e.g.,][]{KL07,DFTKH10,FRD+15}, 
or by a massive planet or embryo that carves a gap in the gas disk \citep[e.g.,][]{PM06,RAWL06,ZNDEH12,MN12,LJM14}.

Interestingly, the snow line has been regraded as a candidate that might naturally produce 
a pressure bump \citep{KL07,DFTKH10,BHD08,DWD13}. 
This idea is based on the assumptions that (i) the dust surface density has a jump there and 
that (ii) the jump in the surface density leads to a jump in the magnetic turbulent viscosity large 
enough to build up a pressure bump via ionization chemistry. 
Although the first assumption is likely to be true qualitatively, the second assumption 
has not yet been validated with magnetohydrodynamic simulations incorporating a realistic ionization model.

By contrast, hydrodynamical simulations have demonstrated 
the viability of particle trapping at the edges of planetary gaps \citep{PM06,ZNDEH12}.
Simulations by \citet{PM06} and \citet{LJM14} show that a planet larger than 
$\sim 20 M_\oplus$ in mass carves a gap that can efficiently trap incoming pebbles.
This suggests that excessive water delivery to terrestrial planetary embryos 
may be avoided if such a massive planet forms prior to the inward migration of the snow line.  
{This possibility has also been pointed out in a recent paper by \citet{MBC+16}.}
Assuming that the snow line moves on a timescale of $\sim 2~{\rm Myr}$ as estimated in Sect.~\ref{sec:gasst},
the standard planet formation from planetesimals is too slow to satisfy this requirement  
unless the disk is massive and the collisional fragmentation of the bodies is negligible \citep{KI02,KTKI10}.
By contrast, planet growth driven by pebble accretion can take place on this timescale 
as demonstrated by \citet{LJ14}.

To conclude, this study has shown that depletion of icy pebbles before the migration of the snow line 
is a possible explanation for the origin of water-devoid terrestrial planets,  
but pebble filtration by a gap-forming planet (in the solar system, Jupiter or Saturn) 
that forms before the snow-line migration might be an equally viable alternative.
The scenario that is favored for the solar system is unclear and needs to be answered in future studies.

\section{Conclusions}\label{sec:summary}
We have studied how icy pebbles drifting from outer regions of protoplanetary disks 
affect the water fraction of terrestrial embryos near the Earth's orbit. 
We numerically solved a simplified version of the coagulation equation 
to calculate the global evolution of the characteristic size and mass flux of drifting icy pebbles.
In contrast to the previous study on pebble accretion by \citet{LJ14}, our model explicitly 
takes  the finite radial extent of a protoplanetary disk into account,  and therefore 
automatically includes the effect that the radial pebble flux diminishes as 
the outermost region of the disk is depleted of icy dust. 
We calculated the rate of pebble accretion by a single embryo following
the analytic expressions by \citet{OK10} and \citet{GIO14}.
Our calculation accounts for the 3D nature of pebble accretion, i.e., 
the reduction of the accretion rate due to vertical pebble diffusion,
in the presence of  strong gas turbulence.
We have predicted how the mass and water content of an initially rocky embryo 
increase with time after icy pebble accretion sets in. 
{The predicted water fraction was compared with the minimum water 
fraction of the present Earth inferred from the ocean mass (0.023 wt\%) and with 
the hypothetical water fraction of the proto-Earth inferred from 
the density deficit of the Earth's outer core (1 wt\%).}

Our key findings are summarized as follows:

\begin{enumerate}
\item 
The evolution of the icy pebble flow 
largely depends on the radial extent of the gas disk $r_{\rm out}$ 
(Sects.~\ref{sec:global} and \ref{sec:1AU}).
In general, radially drifting pebbles form from inside out in protoplanetary disks  
because the timescale of pebble formation roughy scales with the orbital timescale.   
The radial extent of a disk therefore sets the lifetime of the pebble flow in the disk
with smaller $r_{\rm out}$ corresponding to a shorter lifetime.  
Turbulence somewhat accelerates pebble formation in outermost regions,
but its effect is relatively minor when compared to the effect of $r_{\rm out}$.
The radial pebble flux is $\sim 10^2 M_\oplus~{\rm Myr}^{-1}$ at early times,
and starts decaying with time at $t \approx 0.2~{\rm Myr}$ for $r_{\rm out} = 100~{\rm AU}$ 
and at $t \approx 1$ Myr for $r_{\rm out} = 300~{\rm AU}$.
The pebble size observed at 1 AU is a few decimeters at early times, 
and decreases with the pebble flux because a lower particle density generally leads to slower particle growth.   
 
\item
The rate of pebble accretion onto an embryo also decreases with time, 
the details of which depend on $r_{\rm out}$ and on turbulence strength $\alpha$ (Sect.~\ref{sec:Medot}). 
The pebble accretion rate is initially $\sim 1 M_\oplus~{\rm Myr}^{-1}$, 
reflecting the fact that the radial mass flux is $\sim 10^2 M_\oplus~{\rm Myr}^{-1}$ and 
the accretion efficiency of dm-sized particles is $\sim 1~\%$ (Figure~\ref{fig:P}).
The accretion rate roughly scales linearly with the embryo mass, 
and therefore the embryo's water fraction increases with time nearly independent of the embryo mass. 
Turbulence suppresses the accretion rate at late times 
by diffusing small pebbles away from the midplane.
Combining this effect with the acceleration of pebble depletion mentioned above, 
strong turbulence of $\alpha = 10^{-2}$ reduces the pebble flux by 1--2 orders of 
magnitude from the cases with weaker turbulence ($\alpha = 10^{-3}$ and $10^{-4}$).


\item 
Preserving water-devoid embryos at 1 AU generally requires that 
the snow line reaches that location after the radial pebble flux through 
the disk has already decayed to a sufficient extent (Sect.~\ref{sec:evol}).   
In a disk of $r_{\rm out} \la 100~{\rm AU}$ and $\alpha \ga 10^{-3}$,
the fractional water content of the embryos is kept below 
the current Earth's water content based on the ocean mass (0.023~{\rm wt\%})
if the time $t_{\rm start}$ at which the snow lines passes 1 AU 
is longer than $2$--4~{\rm Myr} (smaller $t_{\rm start}$ corresponds to stronger turbulence).
If $r_{\rm out} \geq 300~\AU$ or $\alpha \leq 10^{-4}$, 
the water content greatly exceeds 0.023 wt\% for all $t_{\rm start} \la 5~{\rm Myr}$.  
Keeping the water fraction at a level of $ \la 1~{\rm wt\%}$ is possible
in a $r_{\rm out} = 100~{\rm AU}$ disk if  $t_{\rm start} \ga 0.5$--2 Myr.
Keeping the water fraction at the same level is much more difficult 
in a $r_{\rm out} = 300~{\rm AU}$ disk because of a prolonged pebble flow. 
\end{enumerate}

Our results provide strong constraints on the formation history and environment 
of Earth-forming embryos in the solar nebula
{within the assumption that no mechanism halted the ice pebble flow upstream.}
Overall, our results suggest that the solar nebula must have been relatively compact 
($r_{\rm out} = 100 \rm AU$ or smaller), {so that the pebble flow had decayed at early times.
The formation of terrestrial embryos as dry as the present Earth
was possible  if moderately strong turbulence ($\alpha \ga 10^{-3}$) was present at 1 AU.
However, the latest} magnetohydrodynamical disk models \citep[e.g.,][]{BS13b,LKF14} 
suggest that turbulence is considerably weaker than this requirement in inner regions of protoplanetary disks.\footnote{In addition, turbulence of $\alpha > 10^{-3}$ would inhibit the formation of 
rocky embryos via rocky pebble accretion within the lifetime of protoplanetary disks 
\citep{JMLB15,MLJB15} for the same reason that it prevents excessive water delivery.}
{For $r_{\rm out} = 100~\rm AU$ and $\alpha = 10^{-4}$, 
embryos as dry as the present Earth would not have formed, 
but embryos containing $\la$ 1 wt\% water would have formed 
if the snow line migrated on a timescale of $\ga$ 2 Myr.}
This is one plausible scenario that can explain the origin of our dry Earth
because the snow line in the nebula could indeed have migrated on a similar timescale (Sect.~\ref{sec:gasst}). 

Another possible scenario, which we have not tested in this study, 
is that a massive planet (a gas giant or its core) formed and created a pebble gap 
in the nebula before the snow line reached 1 AU (Sect.~\ref{sec:filter}; {see also \citealt{MBC+16}}).
The scenario favored as the explanation for the dry Earth remains to be studied in future work.

One important caveat of this study is that 
our pebble evolution model greatly simplifies the collisional growth of ice aggregates.
Future models should include the evolution of aggregate porosity \citep{OST07,OTS09,KTOW13a}
since porous aggregates tend to collide more frequently than 
compact equivalents in protoplanetary disks \citep{OTKW12,KTOW13b}.
The global simulation of dust coagulation and porosity evolution by \citet{OTKW12} shows 
that highly fluffy ice aggregates produced in inner ($\la 10~{\rm AU}$) protoplanetary disks 
grow to planetesimal-mass objects with little appreciable drift.  
However, the simulation also shows that fluffy aggregates forming in the outer 
($\ga 10~{\rm AU}$) region drift in until they arrive at the snow line.
The accretion rate of these fluffy aggregates onto inner embryos could be greatly different 
from that of compact aggregates.
Bounding and fragmentation of aggregates are also potentially important, 
but might not be crucial 
given the sticky nature of ice aggregates \citep[see][]{WTS+09,WTS+11,WTO+13,GB15}.

Another important caveat is that it is still a matter of debate 
whether the snow line around a solar-mass star really migrates inward to 1 AU.
The model of \citet{ML12}, which predicts that the snow line never reaches 1 AU in late evolutionary stages,
relies on the idea that weak magnetic turbulence expected at $\sim$ 1 AU
\citep[e.g.,][]{G96,SMUN00} triggers gravitational instability that in turn heats up the disk gas at that location.  
On the other hand, recent magnetohydrodynamical simulations have shown that, even without magnetic turbulence, large-scale (non-turbulent) magnetic fields are still able to provide inner protoplanetary disks 
with high gas accretion rates \citep[e.g.,][]{TS08,BS13b,LKF14,GTNM15}.
A model incorporating these important accretion mechanisms is needed 
to fully understand the evolution of the snow line.

{The results of this study also have important implications for the water content of exoplanets 
lying inside the habitable zone. \citet{MCMP15} recently predicted the water content distribution 
of habitable-zone terrestrial planets assuming that the planets acquire water  
by accreting water-bearing (water content = 5\%) embryos and planetesimals. 
However, our results suggest that a significantly higher amount of water could be delivered to 
habitable-zone planets in the form of icy pebbles from outer orbital radii. 
A significant amount of water does not necessarily make habitable-zone planets habitable
because the presence of land might be required for the emergence of life \citep{DM15}.
In any case, it would be interesting to predict the water content of habitable-zone exoplanets by taking ice pebble accretion into account. 
}

\begin{acknowledgements}
We would like to thank Takanori Sasaki 
for pointing out the importance of pebble accretion in the context of water delivery to the Earth. 
We also thank Tristan Guillot and Chris Ormel for discussions on the modeling of pebble accretion; 
Taishi Nakamoto, Hidenori Genda, Masanobu Kunitomo, Tetsuo Taki, for helpful comments;
{and Sebastiaan Krijt for sharing with us an early version of his paper prior to publication.}
S.\,O. especially thanks Chris Ormel for his very insightful comments 
on the formulation of the single-size approach based on the moment method. 
{Finally, we thank the referee, Michiel Lambrechts, for his prompt and constructive report which 
significantly improved the quality of this paper. }
This work is supported by Grants-in-Aid for Scientific Research (\#23103005, 15H02065) 
from MEXT of Japan. 
\end{acknowledgements}

\appendix
\section{Derivation and justification of the single-size approach}\label{sec:simple}
\subsection{Derivation of Eqs.~\eqref{eq:evol_Sigmad} and \eqref{eq:evol_mstar} from the coagulation equation}
\label{sec:moment}
In this subsection, we derive the single-size equations \eqref{eq:evol_Sigmad} and \eqref{eq:evol_mstar} 
from the coagulation (Smoluchowski) equation. 
We define the size distribution function $n(r,z,m)$ as the particle number density per unit particle 
mass $m$ at orbital radius $r$ and distance from midplane $z$. 
Assuming the balance between vertical sedimentation and turbulent diffusion of the particles,
the particle size distribution can be written as 
$n = (\mathcal{N}/\sqrt{2\pi}h_{\rm d}){\rm exp}(-z^2/2h_{\rm d}^2)$,
where ${\cal N}(r,m)$ is the column number density of dust particles per unit $m$ 
and {$h_{\rm d}(m)$} is the dust scale height. 
The evolution of $\mathcal{N}$ is given by the vertically integrated coagulation equation 
with the drift term \citep{BDH08}
\begin{align}
\frac{\partial \mathcal{N}(r,m)}{\partial t} 
&= \frac{1}{2}\int_{0}^{m} K(r; m',m-m')\mathcal{N}(r,m')\mathcal{N}(r,m-m')dm' 
\notag \\
&\quad - \mathcal{N}(r,m)\int_{0}^{\infty} K(r; m,m')\mathcal{N}(r,m')dm' 
\notag \\
&\quad - \frac{1}{r}\frac{\partial}{\partial r}\left[ rv_{\rm r}(r,m)\mathcal{N}(r,m) \right].
\label{eq:coag_N}
\end{align}
Here, $K$ is the vertically integrated collision rate coefficient given by
\begin{equation}
K(r; m_{1},m_2) = \frac{{\sigma}_{\rm coll}}{2{\pi}h_{\rm d,1}h_{\rm d,2}}\int_{-\infty}^{\infty}
{\Delta}v_{\rm pp}\,
{\exp}\left[
-\frac{z^2}{2}\left(\frac{1}{h_{\rm d,1}^2} + \frac{1}{h_{\rm d,2}^2}\right)\right]dz, \label{eq:coeffK}
\end{equation}
where $h_{\rm d,1}$ and $h_{\rm d,2}$ are the scale heights of the two colliding particles.
{On the right-hand side of Eq.~\eqref{eq:coag_N}, the first term represents 
the gain of ${\cal N}(m)$ by coagulation of two particles of masses $m'$ and $m-m'$,
the second term the loss of ${\cal N}(m)$ by coagulation of a particle of mass $m$ with a particle 
of mass $m'$, and the third term the advection of ${\cal N}(m)$ due to radial drift. }
Because we assume  perfect sticking upon collision, the collisional cross section is simply given by
${\sigma}_{\rm coll} = {\pi}(a_1 + a_2)^2$, where $a_1$ and $a_2$ are the particle radii.

Since we are interested in the mass flow of radially drifting dust particles, 
it is useful to introduce the surface {\it mass} density of dust per unit particle mass
\beq
{\cal S}(r,m) \equiv m{\cal N}(r,m).
\eeq 
Multiplying Eq.~\eqref{eq:coag_N} by $m$, the equation for ${\cal S}$ is obtained as  
\begin{align}
\frac{\partial{\cal S}(m)}{\partial t} &= \frac{m}{2}\int_{0}^{m} K'(m',m-m')
{\cal S}(m'){\cal S}(m-m')dm' 
\notag \\
&\quad - m{\cal S}(m) \int_{0}^{\infty} K'(m,m'){\cal S}(r,m')dm'
\notag \\
&\quad- \frac{1}{r}\frac{\partial}{\partial r} 
\left[  r v_{\rm r}(m){\cal S}(m) 
\right],
\label{eq:coag_S}
\end{align}
where $K'(m,m') \equiv K(m,m')/mm'$ and we have omitted the argument $r$ from the expression
for clarity.  

One important quantity characterizing the mass distribution ${\cal S}$ is the 
so-called peak mass defined by \citep{OS08}
\begin{equation}
m_{\rm p} \equiv \frac{\int m {\cal S}dm}{\int {\cal S} dm}.
\label{eq:mp_def}
\end{equation}
When ${\cal S}$ is a unimodal function of $m$, this quantity is approximately equal to 
the mass at the peak of ${\cal S}$ \citep[see][]{OS08}.   
Another important quantity is of course the total surface density defined by 
\beq
\Sigma_{\rm d} \equiv \int_{0}^{\infty} {\cal S}(m) dm.
\label{eq:Sigmad_def}
\eeq

We now derive the equations that determine the evolution of $\Sigma_{\rm d}$ and $m_{\rm p}$ 
from Eq.~\eqref{eq:coag_S}.
We begin by introducing the $i$-th moment of the surface mass density distribution ${\cal S}$,
\begin{equation}
{\cal M}_i \equiv \int_{0}^{\infty} m^i {\cal S}(m) dm. 
\end{equation}
The quantity ${\cal M}_i$ is equal to the $(i+1)$-th moment of the 
surface number density distribution ${\cal N}$ since ${\cal S} = m{\cal N}$.
It follows from Eqs.~\eqref{eq:mp_def} and \eqref{eq:Sigmad_def}
that the total dust surface density $\Sigma_{\rm d}$ and peak mass $m_{\rm p}$ are related to ${\cal M}_{\rm i}$ as 
\beq
\Sigma_{\rm d}  = {\cal M}_0,
\label{eq:Sigmad_M}
\eeq
\beq
m_{\rm p} = \frac{{\cal M}_1}{{\cal M}_0},
\label{eq:mp_M}
\eeq
respectively.

The equation that determines the evolution of ${\cal M}_i$ can be derived by multiplying Eq.~\eqref{eq:coag_S} 
by $m^i$ and integrating over $m$. 
After some algebra, the result can be simplified as 
\citep[see, e.g.,][but note the they define the moment in terms of the number density]{EC08,OS08}
\begin{align}
\frac{\partial{\cal M}_i}{\partial t} 
&= \frac{1}{2}\int_0^\infty dm\, dm' \, K'(m,m'){\cal S}(m){\cal S}(m')
\notag \\
&\qquad \times [(m+m')^{i+1}-(m^{i+1}+m'^{i+1})]
\notag \\
&\quad - \frac{1}{r}\frac{\partial}{\partial r} 
\left( r \bracket{m^i{v}_{\rm r}}{{\Sigma}}_{\rm d}
\right),
\label{eq:evol_Mp}
\end{align}
where 
\begin{equation}
\bracket{m^i v_{\rm r}} \equiv \frac{1}{{\Sigma}_{\rm d}}\int_{0}^{\infty} m^i v_{\rm r}(m){\cal S}(r,m)dm.
\end{equation}
For $i = 0$, Eq.~\eqref{eq:evol_Mp} has the simple form
\beq
\frac{\partial \Sigma_{\rm d}}{\partial t} 
= \frac{1}{r}\frac{\partial}{\partial r} \left(
r \bracket{{v}_{\rm r}}{{\Sigma}}_{\rm d}
\right).
\label{eq:evol_M0} 
\eeq
The coagulation terms (the first and second terms) in Eq.~\eqref{eq:evol_Mp} have
canceled out, as it should be since ${\Sigma}_{\rm d}$ is 
a conserved quantity in the absence of advection and diffusion. 
Since ${\cal M}_1 = m_{\rm p}\Sigma_{\rm d}$, the equation for $i=1$ can be written as
\begin{align}
\frac{\partial(m_{\rm p}\Sigma_{\rm d})}{\partial t} 
&= \int_0^\infty  dm\, dm' \, K(m,m'){\cal S}(m){\cal S}(m')
\notag \\
&- \frac{1}{r}\frac{\pd}{\pd r}\left( r \bracket{m v_{\rm r}} \Sigma_{\rm d} \right).
\label{eq:evol_M1} 
\end{align}

The right-hand sides of Eq.~\eqref{eq:evol_M0} and \eqref{eq:evol_M1} are not closed 
with respect to $\Sigma_{\rm d}$ and $m_{\rm p}$ because of the presence of the $\bracket{m^i v_{\rm r}} $ terms. 
To derive approximate but {\it closed} equations for $\Sigma_{\rm d}$ and $m_{\rm p}$, 
we assume that ${\cal S}$ is narrowly peaked at $m \approx m_{\rm p}$. Specifically, we assume
\begin{equation}
{\cal S}(m) \approx {\Sigma}_{\rm d}{\delta}(m - m_{\rm p}), 
\label{eq:sigmadel}
\end{equation}
where the normalization is determined by Eq.~\eqref{eq:Sigmad_M}.
If we use \eqref{eq:sigmadel}, the term $\bracket{m^i v_{\rm r}}$ can now be replaced by
$m_{\rm p}^i v_{\rm r}(m_{\rm p})$, and hence Eq.~\eqref{eq:evol_M0} immediately reduces to Eq.~\eqref{eq:evol_Sigmad}
in the main text. 
Equation~\eqref{eq:evol_M1}  reduces to  
\beq
\frac{\partial(m_{\rm p}\Sigma_{\rm d})}{\partial t} 
\approx K_{\rm pp} \Sigma_{\rm d}^2 
 - \frac{1}{r}\frac{\partial}{\partial r} \left( r 
 m_{\rm p}{v}_{\rm r}(m_{\rm p}){{\Sigma}}_{\rm d}
\right),
\label{eq:evol_M2}
\eeq
where 
\beq
K_{\rm pp} \equiv K(m_{\rm p},m_{\rm p}) = \frac{2a^2}{h_{\rm d}(m_{\rm p})^2}
\int_{-\infty}^\infty \Delta v\,\exp\left( -\frac{z^2}{h_{\rm d}(m_{\rm p})^2}\right )dz
\label{eq:Kstar}
\eeq 
and $a$ is the radius of peak-mass particles.
If $\Delta v$ can be taken outside the vertical integration,  
we have $K^* = 2\sqrt{\pi} a^2\Delta v/h_{\rm d}$.
Substituting this expression into Eq.~\eqref{eq:evol_M2} and 
combining with Eq.~\eqref{eq:evol_Sigmad}, 
we obtain Eq.~\eqref{eq:evol_mstar} in the main text.

\subsection{Comparison with full-size calculations}\label{sec:comparison}
To test the validity of the simple size approach, we attempt to reproduce the result of 
a full size calculation by \citet{OTKW12}. 
They calculated the evolution of the full size distribution with and without porosity evolution. 
We select the result of the compact aggregation model 
where the particle internal density is fixed to $1.4~{\rm g~cm^{-3}}$. 
In accordance with \citet{OTKW12}, we assume the optically thin MMSN 
with temperature profile $T = 280(r/1\AU)^{-1/2}~{\rm K}$ and compute dust evolution 
in the region $3~{\rm AU} < r < 150~{\rm AU}$. 
The particle collision velocity $\Delta v_{\rm pp}$ 
is given by Eq.~\eqref{eq:vpp} in Sect.~\ref{sec:velocity}.
As we explained there, we introduce a free parameter $\epsilon$ to $\Delta v_{\rm pp}$ 
to account for the effect of a finite size dispersion.  
Below we consider three choices $\epsilon = 0$, 0.5, and 1.

\begin{figure} 
\centering
\includegraphics[width=9cm]{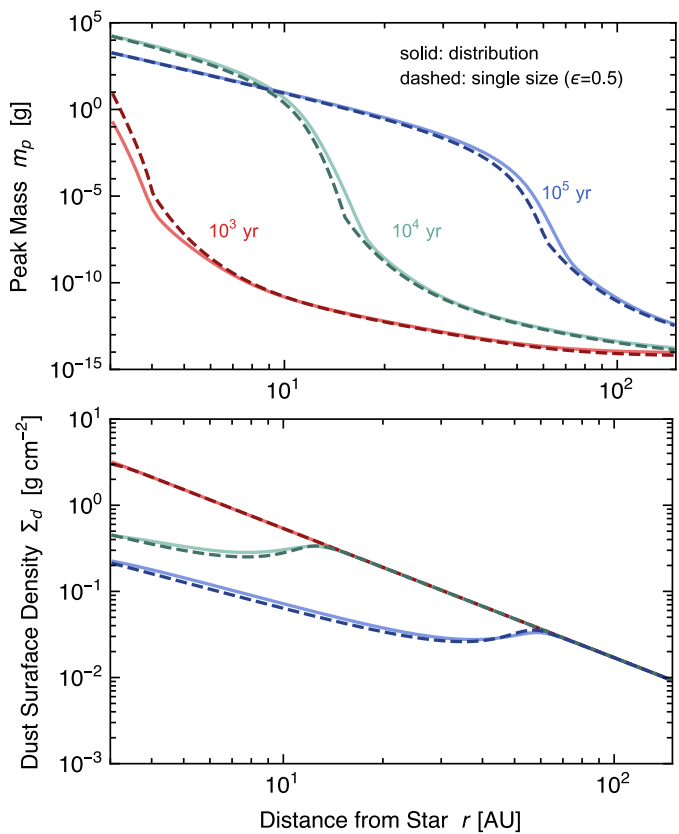}
\caption{Comparison between single- and full-size coagulation calculations. 
The solid lines show the snapshots of the peak mass $m_{\rm p}$ (Eq.~\eqref{eq:mp_M}; upper panel)
and total dust surface density $\Sigma_{\rm d}$ (Eq.~\eqref{eq:Sigmad_M}; lower panel)
at different times as a function of orbital radius $r$ obtained from the full coagulation 
simulation for the compact aggregation model by \citet[][see also their Fig.~2]{OTKW12}.
The dashed line shows our reproduction using the single-size approach (Eq.~\eqref{eq:evol_Sigmad} 
for $\Sigma_{\rm d}$ and Eq.~\eqref{eq:evol_mstar} for $m_{\rm p}$) with $\epsilon = 0.5$
(see Sect.~\ref{sec:velocity} for the definition of $\epsilon$). 
}
\label{fig:21}
\end{figure}
Figure \ref{fig:21} shows the radial distribution of the particle peak mass $m_{\rm p}$ 
and total dust surface density $\Sigma_{\rm d}$ at different times  
obtained from the full size calculation by \citet[][see their Fig.~2 
for the corresponding snapshots of the size distribution]{OTKW12}.
These are directly obtained from the data of the full size distribution 
together with the definitions of $m_{\rm p}$ and $\Sigma_{\rm d}$, 
Eqs.~\eqref{eq:Sigmad_M} and~\eqref{eq:mp_M}.
We find that the single-size calculation with $\epsilon = 0.5$ 
reproduces these results with reasonably good accuracy (see Figure \ref{fig:21}).  
The agreement is particularly good for drifting pebbles (e.g., for $t=10^5~{\rm yr}$, 
particles at $\la$ 60~AU) whose mass are determined by 
the balance between radial drift and local coagulation.  

\begin{figure} 
\centering
\includegraphics[width=9cm]{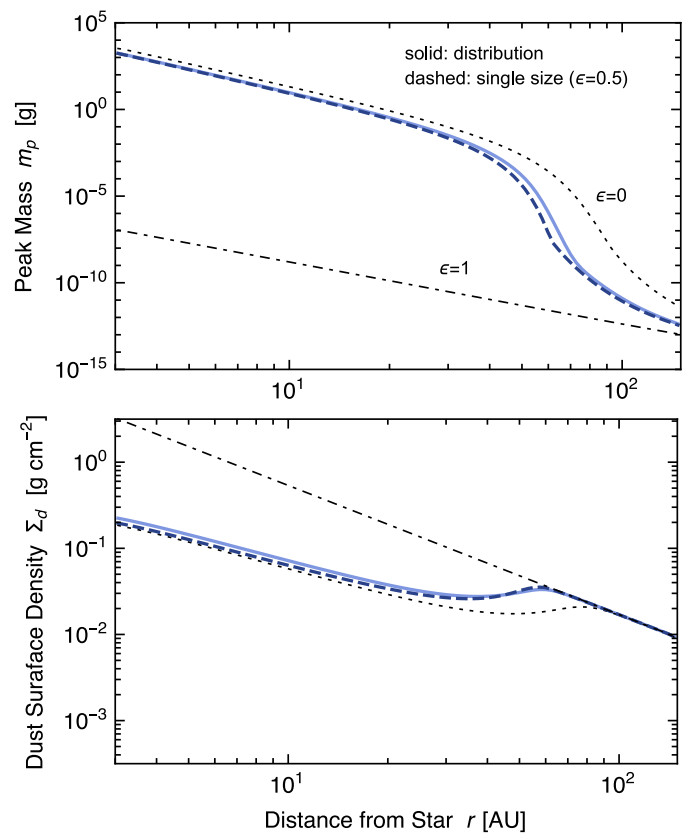}
\caption{Comparison between the full-size simulation \citep{OTKW12} 
and single-size simulations using different values of $\epsilon$.
The solid lines show the snapshots of the peak mass $m_{\rm p}$ (Eq.~\eqref{eq:mp_M}; upper panel)
and total dust surface density $\Sigma_{\rm d}$ (Eq.~\eqref{eq:Sigmad_M}; lower panel)
at $10^5~{\rm yr}$, while the dotted, dashed, and dash-dotted lines show the reproductions 
with $\epsilon = 0$, 0.5, and 1, respectively. 
The comparison shows that the single-size approach best reproduces the result of the 
full-size simulation when $\epsilon$ is chosen to be $0.5$. 
}
\label{fig:22}
\end{figure}
Figure \ref{fig:22} demonstrates the importance of taking into account the effect 
of {a} finite size dispersion in evaluating $\Delta v_{\rm pp}$. 
Here we compare the snapshots of the single-size calculations 
with different values of $\epsilon$ at $t = 10^5~{\rm yr}$.
We can see that the single-size calculation significantly underestimates 
the rate of particle evolution if the effect of size dispersion is ignored, i.e., $\epsilon = 1$.
The reason is that equal-sized particles have vanishing non-Brownian relative velocities
when they are so small (e.g., $a \la 10^{-3}~{\rm cm}$ at 100 AU) that their 
stopping time is shorter than the turnover time of the smallest turbulent eddies \citep[see, e.g.,][]{OC07}.
We find that the opposite limit, $\epsilon = 0$, gives a much better agreement 
and the intermediate choice, $\epsilon = 0.5$, gives the best match to the full solution.

\bibliographystyle{aa}
\bibliography{myrefs_160211}

\end{document}